%% file: MUO-21-001_temp.tex
\pdfoutput=1
\documentclass[11pt,twoside,a4paper,cmspaper,final,collab]{cms-tdr}
\def\svnVersion{1e08ab8}\def\svnDate{2024/09/06}\def\cmsCernNoTag{CERN-EP-2024-040}\def\cmsCernDate{\today}\def\cmsMessage{Published in the Journal of Instrumentation as \href{http://dx.doi.org/10.1088/1748-0221/19/09/P09012}{\doi{10.1088/1748-0221/19/09/P09012}.}}
\begin{document}\cmsNoteHeader{MUO-21-001}

\newlength\cmsFigWidth
\ifthenelse{\boolean{cms@external}}{\setlength\cmsFigWidth{0.8\columnwidth}}{\setlength\cmsFigWidth{0.48\textwidth}}
\newlength\cmsFigWidthWide
\ifthenelse{\boolean{cms@external}}{\setlength\cmsFigWidthWide{0.8\columnwidth}}{\setlength\cmsFigWidthWide{0.6\textwidth}}
\newlength\cmsFigWidthWideSchema
\ifthenelse{\boolean{cms@external}}{\setlength\cmsFigWidthWideSchema{0.95\columnwidth}}{\setlength\cmsFigWidthWideSchema{0.75\textwidth}}
\newlength\cmsFigWidthVeryWide
\ifthenelse{\boolean{cms@external}}{\setlength\cmsFigWidthVeryWide{1.0\textwidth}}{\setlength\cmsFigWidthVeryWide{1.0\textwidth}}

\ifthenelse{\boolean{cms@external}}{\providecommand{\cmsLeft}{top\xspace}}{\providecommand{\cmsLeft}{left\xspace}}
\ifthenelse{\boolean{cms@external}}{\providecommand{\cmsRight}{bottom\xspace}}{\providecommand{\cmsRight}{right\xspace}} 
\ifthenelse{\boolean{cms@external}}{\providecommand{\cmsLLeft}{Top\xspace}}{\providecommand{\cmsLLeft}{Left\xspace}}
\ifthenelse{\boolean{cms@external}}{\providecommand{\cmsRRight}{Bottom\xspace}}{\providecommand{\cmsRRight}{Right\xspace}} 
\ifthenelse{\boolean{cms@external}}{\providecommand{\Apdx}{}}{\providecommand{\Apdx}{Appendix}} 

\newlength\cmsTabSkip\setlength{\cmsTabSkip}{1ex}
\ifthenelse{\boolean{cms@external}}{\providecommand{\cmsTable}[1]{#1}}{\providecommand{\cmsTable}[1]{\resizebox{\textwidth}{!}{#1}}}
\ifthenelse{\boolean{cms@external}}{\providecommand{\cmsTableSmall}[1]{\resizebox{\columnwidth}{!}{#1}}}{\providecommand{\cmsTableSmall}{}}

\newcommand{\EPOS} {{\textsc{epos}}\xspace}
\newcommand{\sqrts}{\ensuremath{\sqrt{s}}\xspace}
\newcommand{\pp}{\ensuremath{\Pp\Pp}\xspace}
\newcommand{\PbPb}{\ensuremath{\mathrm{PbPb}}\xspace}
\newcommand{\pPb}{\ensuremath{\Pp{}\mathrm{Pb}}\xspace}
\newcommand{\Ntracks}{\ensuremath{N_{\mathrm{tracks}}}\xspace}
\newcommand{\ereco}{\ensuremath{\epsilon_\text{reco\textbar track}}\xspace}
\newcommand{\eID}{\ensuremath{\epsilon_\text{ID\textbar reco}}\xspace}
\newcommand{\etrig}{\ensuremath{\epsilon_\text{trig\textbar ID}}\xspace}
\newcommand{\etrack}{\ensuremath{\epsilon_\text{track}}\xspace}
\newcommand{\dimuonRap}{\ensuremath{y^{\PGm\PGm}}\xspace}
\newcommand{\Dz}{\ensuremath{D_{\mathrm{z}}}\xspace}
\newcommand{\Dxy}{\ensuremath{D_{\mathrm{xy}}}\xspace}
\newcommand{\mPDG}{\ensuremath{m_{\textrm{PDG}}}\xspace}
\newcommand{\mfit}{\ensuremath{m_{\text{fit}}}\xspace}
\newcommand{\sigmaCB}{\ensuremath{\sigma_{\textrm{CB}}}\xspace}

\cmsNoteHeader{MUO-21-001}
\title{Performance of CMS muon reconstruction from proton-proton to heavy ion collisions}

\date{\today}

\abstract{The performance of muon tracking, identification, triggering, momentum resolution, and momentum scale has been studied with the CMS detector at the LHC using data collected at $\sqrtsNN=5.02\TeV$ in proton-proton (\pp) and lead-lead (\PbPb) collisions in 2017 and 2018, respectively, and at $\sqrtsNN = 8.16\TeV$ in proton-lead (\pPb) collisions in 2016. Muon efficiencies, momentum resolutions, and momentum scales are compared by focusing on how the muon reconstruction performance varies from relatively small occupancy \pp collisions to the larger occupancies of \pPb collisions and, finally, to the highest track multiplicity \PbPb collisions. We find the efficiencies of muon tracking, identification, and triggering  to be above 90\% throughout most of the track multiplicity range. The momentum resolution and scale are unaffected by the detector occupancy. The excellent muon reconstruction of the CMS detector enables precision studies across all available collision systems.
}

\hypersetup{
pdfauthor={CMS Collaboration},
pdftitle={Performance of CMS muon reconstruction in heavy ion collisions},
pdfsubject={CMS, muons},
pdfkeywords={CMS, muons, heavy ion collisions}}

\maketitle 

\section{Introduction}

An estimate of the capability of a detector to reconstruct and identify particles of interest is essential to ensure the reliability of the data and the precision of measurements. A crucial measure of the performance of both the detector and its reconstruction algorithms is the efficiency, which is a required correction for any measurement of cross sections 
or particle yields. Efficiency is defined as a fraction of a given species of particles within the detector acceptance that are successfully reconstructed with a given set of selection criteria. The efficiency could be estimated using Monte Carlo (MC) simulations, where the particles present in the event are known a priori. However, estimates from MC simulations assume an accurate description of a detector and its interaction processes, which may have characteristics that are challenging to quantify. Rather than relying solely on detector simulations, data-driven approaches for estimating efficiencies have been developed. We use the data-driven tag-and-probe method that has been successfully used in various analyses in proton-proton (\pp), proton-lead (\pPb), and lead-lead (\PbPb) collisions performed by the CMS detector at the CERN Large Hadron Collider (LHC)~\cite{WZ_pp_2011, W_pPb_2015, Chatrchyan_2012}. Performance studies of the CMS detector have been carried out since the beginning of LHC operations in \pp collisions~\cite{ppMuon_2012, MUO-17-001}, and we now extend these studies to \pPb and \PbPb collisions. 
 
We use heavy ion collisions to gain insight into the collective thermodynamic properties of cold and hot QCD matter, including the color-deconfined quark-gluon plasma (QGP)~\cite{Busza:2018rrf}. Resonances that decay into muons (\eg, \JPsi, \PgU, \Z) provide information about the QGP, because its presence has measurable effects on particle yields and distributions. Muon measurements are more challenging in the high-occupancy environments of heavy ion collisions. One quantifiable indicator of occupancy is the charged particle multiplicity of the event, which in \PbPb collisions is highly correlated with the collision impact parameter (distance between the centers of the two nuclei in the plane perpendicular to the beam line). For the collisions with near-zero impact parameter (central events), the multiplicities can be hundreds of times larger than those observed in a single \pp collision~\cite{CMS:2011aqh}. These central events result in extremely high multiplicity per unit area, especially in the innermost CMS subdetectors used for tracking. Consequently, any degradation in tracking performance is likely to affect the muon measurements, despite the fact that the occupancies in the outermost CMS muon subdetectors remain low. 

In this paper we report on the performance of muon reconstruction, identification (ID), triggering, dimuon-mass resolution, and dimuon-mass scale. The reference physical processes are the decays of \Z and \JPsi resonances, whose decay muons cover a wide kinematic range, thus providing a good data sample for measuring the muon performance in the CMS detector. The efficiencies are estimated using a tag-and-probe method, which is described in detail in Section~\ref{sec:extraction}. The efficiencies are reported in Sections~\ref{sec:trackingEff}--\ref{sec:Jpsi}; the mass resolution and scale are presented in Section~\ref{sec:scaleAndRes}. We summarize the paper in Section~\ref{sec:summary}.

\section{CMS detector}

\begin{figure}[htb]
\centering
\includegraphics[width=\cmsFigWidthWideSchema]{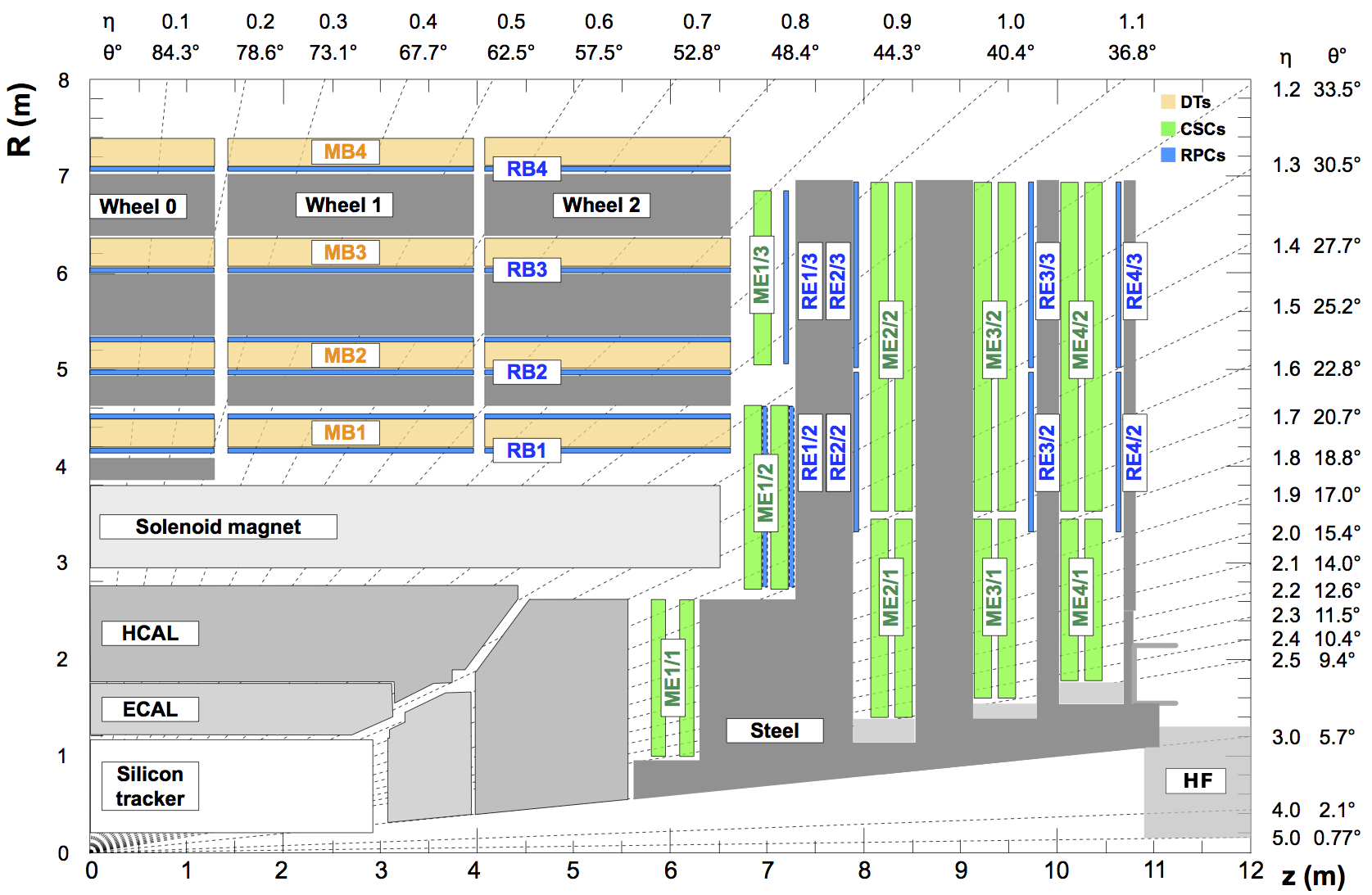}
\caption{Longitudinal layout of one quadrant of the CMS detector. The drawing shows four DT stations in the muon barrel (MB1--MB4, yellow), four CSC stations in the muon endcap (ME1--ME4, green) and the RPC stations (RB1--RB4 and RE1--RE4, blue).}
\label{fig:MuonSystem}
\end{figure}

A detailed description of the CMS detector is reported in Ref.~\cite{CMS_2008}. A schematic view of the detector is shown in Fig.~\ref{fig:MuonSystem}. Muon reconstruction is performed using the all-silicon inner tracker at the center of the detector immersed in a 3.8\unit{T} solenoidal magnetic field, and with up to four stations of gas-ionization muon detectors installed outside the solenoid and sandwiched between the layers of the steel flux-return yoke. The inner tracker is composed of a pixel detector and a silicon strip tracker, and measures charged-particle trajectories in the pseudorapidity range $\abs{\eta}<2.5$. The muon system covers the pseudorapidity region $\abs{\eta}<2.4$ and performs three main tasks: triggering on muons, identifying muons, and improving the momentum measurement and charge determination of high transverse momentum (\pt) muons. Drift tube (DT) detectors and cathode strip chambers (CSC) detect muons in the $\eta$ regions of $\abs{\eta}<1.2$ and $0.9<\abs{\eta}<2.4$, respectively, and are complemented by a system of resistive-plate chambers (RPC) covering the range of $\abs{\eta}<1.9$. The coverage of the DT and CSC detectors defines three regions, referred to as barrel ($\abs{\eta}<0.9$), overlap ($0.9<\abs{\eta}<1.2$), and endcap ($1.2<\abs{\eta}<2.4$) regions. The electromagnetic calorimeter (ECAL) and the hadron calorimeter (HCAL), divided into barrel and endcap sections, are also shown in Fig.~\ref{fig:MuonSystem}. The hadron forward (HF) calorimeters extend the absolute pseudorapidity range of the HCAL to 5.0, and are commonly used in trigger algorithms and estimates of the collision impact parameter.

Events of interest are acquired using a two-tiered trigger system~\cite{Khachatryan:2016bia}. The first level (L1), composed of custom hardware processors, uses information from the calorimeters and muon detectors to select events at a rate of around 100\unit{kHz} for \pp and \pPb collisions (30\unit{kHz} for \PbPb collisions) within a fixed latency of 4\mus~\cite{Sirunyan:2020zal}. The second level, known as the high-level trigger (HLT), consists of a farm of processors running a version of the full event reconstruction software optimized for fast processing. The HLT algorithms, which are seeded by the L1 triggers, further reduce the rate of data collection to around 24\unit{kHz} for the \pp collisions used in this study. The corresponding HLT rate is up to 20\unit{kHz} for \pPb collisions and 7\unit{kHz} for \PbPb collisions.

\section{Data and Monte Carlo samples}
\label{sec:DataMC}

The data used in this paper were taken during the dedicated heavy ion run periods of Run 2 between 2016 and 2018. The events were recorded from \pPb collisions in 2016 at a nucleon-nucleon center-of-mass energy of $\sqrtsNN=8.16\TeV$, from \pp collisions in 2017 at $\sqrts = 5.02\TeV$, and from \PbPb collisions in 2018 at $\sqrtsNN=5.02\TeV$. In addition, a smaller \pPb data sample at $\sqrtsNN=5.02\TeV$ was collected in 2016, which allowed us to quantify the difference between 5.02\TeV and 8.16\TeV collisions. All studied events are from hadronic collisions.

Results from the data are compared with those obtained from MC samples, in which the simulation includes the full CMS detector response using the \GEANTfour package~\cite{Agostinelli:2002hh}. 
The software used to generate the nucleon-nucleon interactions that produce the particle of interest depends on the process and collision system: for \JPsi mesons we use {\PYTHIA}8 (v2.3.0)~\cite{Sjostrand:2014zea}; for \Z bosons in \PbPb, \pPb, and \pp collisions we use \MGvATNLO (v2.4.2)~\cite{Alwall:2014hca}, \POWHEG v2~\cite{Nason:2004rx,Frixione:2007vw,Alioli:2010xd,Alioli:2008gx}, and {\PYTHIA}8, respectively. Differences in the details of the simulation of the hard process have a negligible impact on the studies presented in this paper. In the case of the \pPb and \PbPb MC samples, the hard-scattering nucleon-nucleon events were embedded into an underlying heavy ion collision that was simulated with either \EPOS\ \textsc{lhc} (v3400)~\cite{Pierog:2013ria} ({\pPb}) or \HYDJET v1.9~\cite{Lokhtin:2005px} ({\PbPb}). After the detailed simulation of the CMS detector response with \GEANTfour, MC events were processed through the trigger emulation and reconstructed by the same algorithms that were used for the collision data sample for a given year and collision system.

\subsection{Possible effects of detector and alignment changes}
\label{subsec:alignment}

The reconstruction performance is influenced not only by varying the collision system from \pp to \PbPb, but also by the changes in the detector during the years when the data were collected. The most significant addition affecting the data was the pixel-detector upgrade after the \pPb data-taking period at the end of 2016. As part of the upgrade, an extra layer was added to the pixel detector in the midrapidity region, and an additional disk was added to both sides of the tracker~\cite{CMS:TrackerPhase1}.

The alignment conditions in the pixel and silicon-strip tracker detector differ across the data sets as well. The \pPb and \PbPb samples considered here were reconstructed using the alignment configurations that were derived during data taking. In contrast, the \pp sample takes advantage of the improved end-of-year reconstruction, in which the alignment constants are rederived using the data collected during the entire year. Two examples of measurements that could be sensitive to the alignment conditions are those of the mass scale and of the mass resolution (Section \ref{sec:scaleAndRes}). The details of the alignment procedure and the effects of the alignment conditions on the \Z boson mass reconstructed via muons were studied in Ref.~\cite{CMS:TRK-20-001}. Although the \Z boson decays are used in the alignment procedure, its mass is not fixed to a specific value. In particular, the alignment parameters obtained for a given system are identical irrespective of the multiplicity of the collision. Hence, the \Z boson remains a useful probe into differences in resolution and scale. It was verified in one of the studies discussed in Ref.~\cite{CMS:TRK-20-001} that the difference in the reconstructed \Z boson mass between the alignment conditions derived during data taking and those derived at the end of the year was ${\lesssim}0.1\%$. Based on the small size of the effect, we conclude that the alignment differences do not significantly affect the comparisons among the \pp, \pPb, and \PbPb data sets.

\section{Muons in the heavy ion environment}

\begin{figure}[htb!]
\centering
\includegraphics[width=\cmsFigWidth]{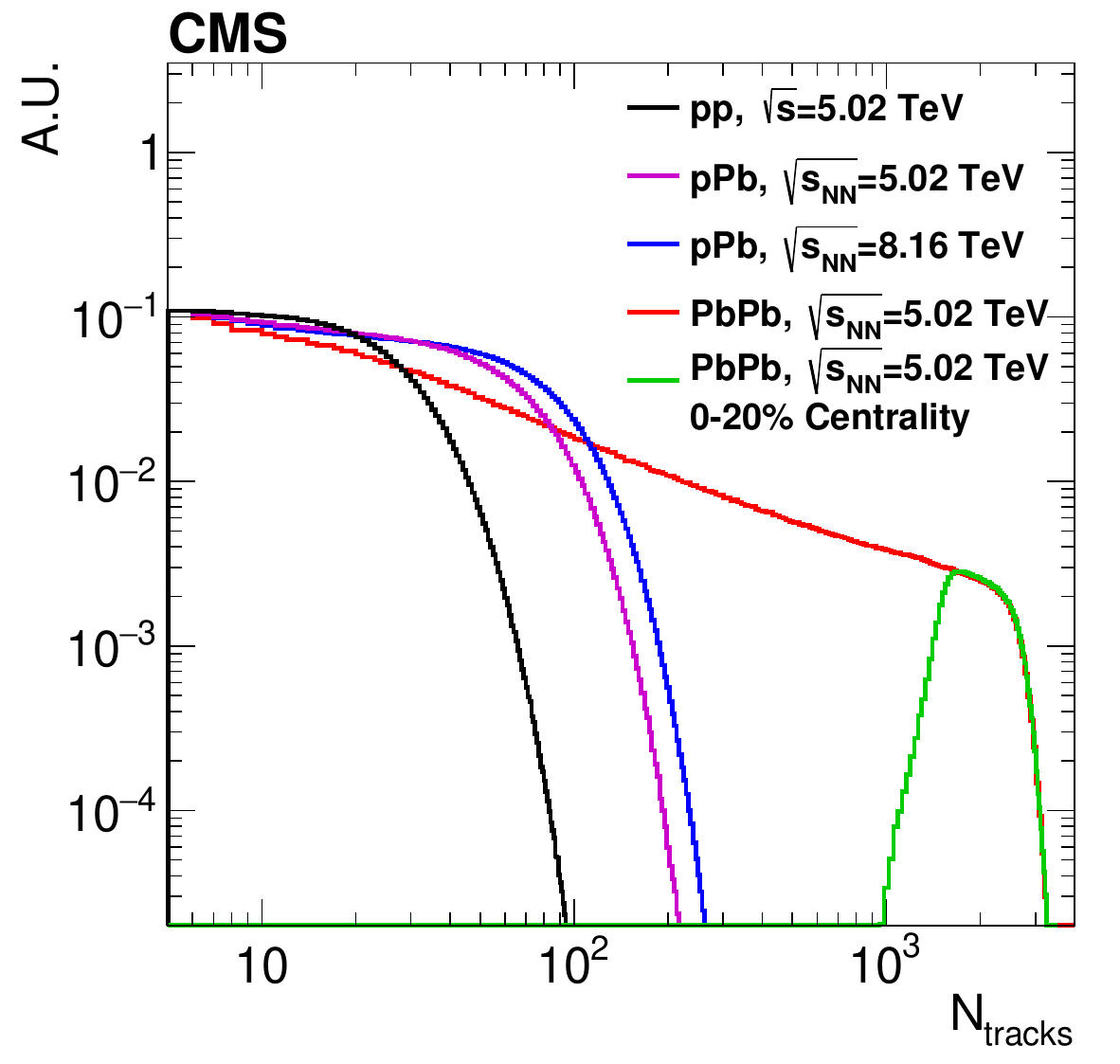}
\includegraphics[width=\cmsFigWidth]{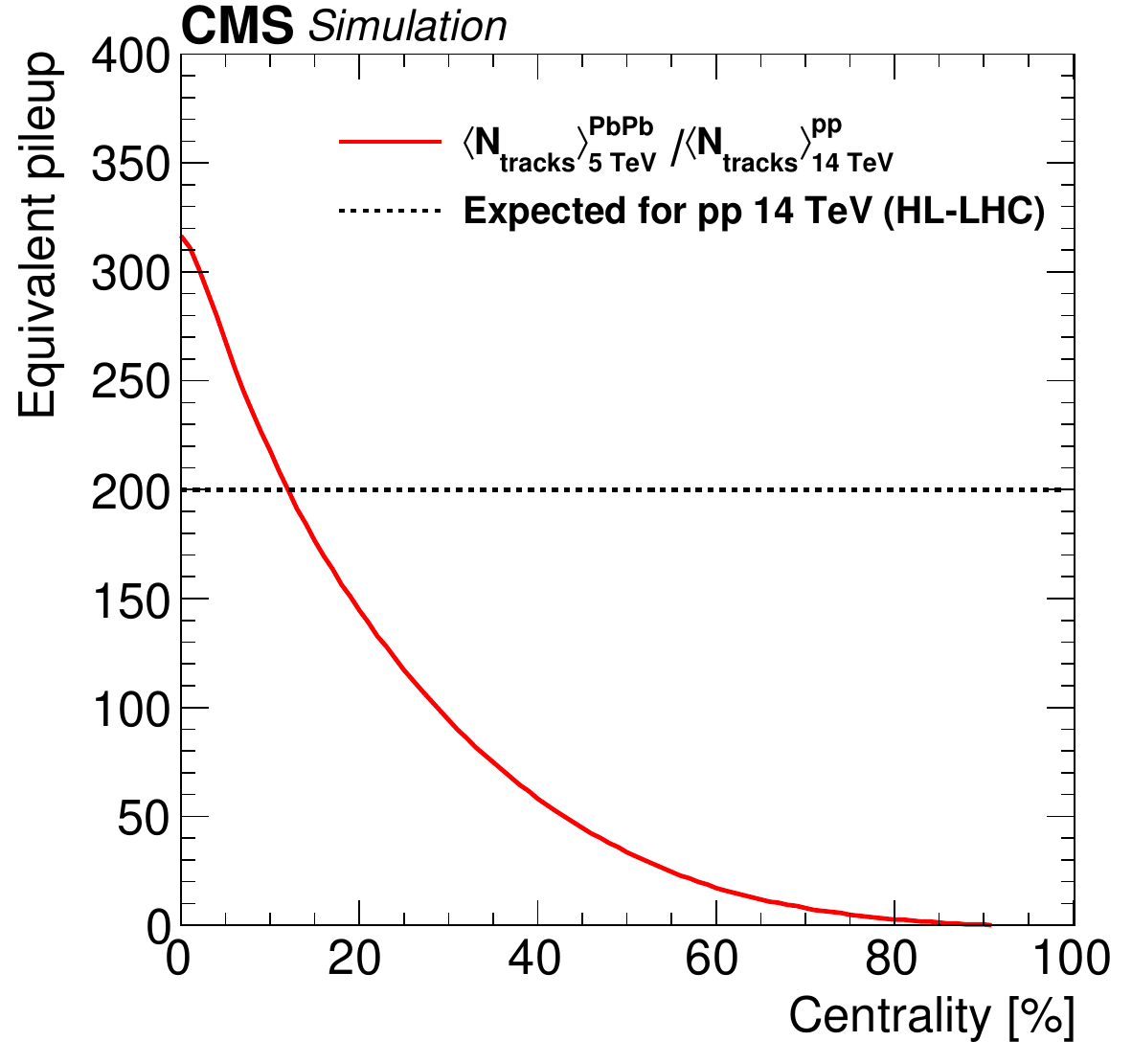}
   \caption{\cmsLLeft: Distribution of \Ntracks in \pp collisions at $\sqrts=5.02\TeV$ (black); \pPb collisions at $\sqrtsNN=5.02\TeV$ (magenta) and at 8.16\TeV (blue); and \PbPb collisions at $\sqrtsNN=5.02\TeV$ (red). The distribution of \Ntracks in the 0--20\%  most central \PbPb collisions is shown in green. The values are not corrected for efficiency. \cmsRRight: Translation of the centrality in \PbPb collisions at $\sqrtsNN=5.02\TeV$ into an equivalent pileup in \pp collisions at $\sqrts=14\TeV$.}
\label{fig:NtracksDists}
\end{figure}

The main difference among the collision systems that affects the detector performance is the number of charged particles produced per collision. The variable \Ntracks, which is obtained by counting tracks in the pseudorapidity range $\abs{\eta}<2.4$, characterizes the total charged particle multiplicity. The distributions of \Ntracks in \pp, \pPb, and \PbPb collision events are shown in the \cmsLeft panel of Fig.~\ref{fig:NtracksDists}. In \PbPb collisions, events with a large number of tracks correspond to the most central collisions (large nucleus-nucleus overlap, small impact parameter), whereas a small number of tracks indicates a peripheral collision (small nucleus-nucleus overlap, large impact parameter). Centrality is a variable used to characterize the amount of overlap between colliding nuclei. The centrality bins are given in percentile ranges of the total inelastic hadronic cross section, calculated from the energy deposited in the HF calorimeters, such that 0\% centrality corresponds to the most central collisions and 100\% corresponds to the most peripheral (see Ref.~\cite{dijets_2011}, Fig.~3, where the centrality definition used in CMS is discussed). To illustrate the highest multiplicities achievable in \PbPb collisions, the distribution of \Ntracks in the 0--20\% most central \PbPb collisions is shown in green. Similarly, larger multiplicities than those of \pp collisions are seen in the \pPb system. We also compare the \Ntracks distributions for \pPb collisions at $\sqrtsNN=5.02\TeV$ and at 8.16\TeV. We see that an increase of 63\% in $\sqrtsNN$ results, on average, in the creation of 24\% more tracks. 

\begin{table}[hbtp]
\centering
\topcaption{Average number of reconstructed tracks, $\langle \Ntracks \rangle$, in our data sets. The values are not corrected for pileup nor efficiency. Two columns are shown for \PbPb collisions. The left column shows the values that are obtained from a minimum-bias data set, and correspond to distributions shown in the \cmsLeft panel of Fig.~\ref{fig:NtracksDists}. The right column shows the values in muon-triggered data, and correspond to the location of points shown in the performance plots of this paper. The values for \pPb are for illustrative purposes only; \pPb results are binned directly in \Ntracks. Details of the centrality conversion to $\langle \Ntracks \rangle$ are provided in \Apdx~\ref{Appendix:CentralityToNtracks}. }
\cmsTableSmall{ 
\begin{tabular}{l cc c}
&  & \multicolumn{2}{c}{$\langle \Ntracks \rangle$ }\\ \cline{3-4}\\[-1.8ex]
{System} & {Centrality range}& minimum-bias & muon-triggered\\
\hline
\multirow{6}{*}{\PbPb, $\sqrtsNN = 5.02\TeV$} & 0--10\% & 2345 & 2397 \\
& 10--20\% & 1721 & 1782 \\
& 20--30\% & 1224 & 1271 \\
& 30--40\% & 819 & 865 \\
& 40--50\% & 503 & 536 \\
& 50--60\% & 279 & 299 \\
& 60--100\% & 33 & 112 \\[\cmsTabSkip]
   \pPb, $\sqrtsNN = 5.02\TeV$ &   & 39 &   \\
   \pPb, $\sqrtsNN = 8.16\TeV$ &   & 48 &   \\[\cmsTabSkip]
   \pp, $\sqrts = 5.02\TeV$ &   & 19 &   \\
\end{tabular}
}
\label{tab:centralityToNtracks}
\end{table}

The \pp collisions involve only two participating nucleons. However, pileup (mean number of collisions per bunch crossing) is significant in the \pp environment due to the high instantaneous luminosities, and this affects the shape and extent of the \Ntracks distribution. Even with pileup, the \pp collisions analyzed in this paper generate far fewer tracks than \PbPb and \pPb collisions. The average number of reconstructed tracks in a given event, $\langle \Ntracks \rangle$, in \PbPb collisions at $\sqrtsNN=5.02\TeV$ corresponding to each centrality range is tabulated in Table~\ref{tab:centralityToNtracks}, and the procedure for obtaining the $\langle \Ntracks \rangle$ is described in \Apdx~\ref{Appendix:CentralityToNtracks}. The pileup for each collision system is shown in Table~\ref{tab:collrate}, which also shows other important characteristics of the collisions discussed in this paper. The pileup in \PbPb collisions is negligible, but must be taken into account in \pp collisions. The multiplicities quoted in Table~\ref{tab:centralityToNtracks} include pileup present in the event. As seen in the table, the 0--10\% most central \PbPb collisions have an average number of tracks $\langle \Ntracks \rangle_{0\text{--}10\%}=2345$. Taking the pileup into account (3 collisions per bunch crossing, as noted in Table~\ref{tab:collrate}), a single \pp collision at $\sqrts=5.02\TeV$ has an average multiplicity of $\langle \Ntracks \rangle \approx 19/3 \approx 6$. This makes the multiplicity in each 0--10\% central \PbPb event roughly equivalent to $2345/6 \approx 400$ \pp collisions occurring simultaneously. The equivalent pileup, $\langle \Ntracks \rangle_{\PbPb}/\langle \Ntracks \rangle_{\pp}$, for a given  \PbPb collision centrality is plotted in the \cmsRight panel of Fig.~\ref{fig:NtracksDists} as a red curve. The calculation uses a simulation of \pp collisions at the highest planned energy for the LHC, 14\TeV, instead of 5.02\TeV. The increase in multiplicity due to this greater \pp energy results in a smaller equivalent pileup for the most central \PbPb events of ${\approx}300$. The expected pileup for \pp collisions in the high-luminosity phase is 200, shown as a black dashed line. Therefore, the central \PbPb collisions correspond to charged particle multiplicities that exceed those expected even in the high-luminosity phase of the LHC. 

\begin{table}[hbtp]
\centering
\topcaption{ Characteristics of selected collisions collected during LHC Run~2. More information on the corresponding luminosity measurements can be found in Refs.~\cite{CMS:LUM-17-003,CMS:LUM-17-002, CMS:LUM-19-001, CMS:LUM-18-001}. }
\cmsTableSmall{ 
\begin{tabular}{lcccc}
    System & \pPb & \pp & \PbPb \\
    \hline
    Year & 2016 & 2017 & 2018 \\
    $\sqrtsNN$ ({\TeVns}) & 8.16 & 5.02 & 5.02 \\
    Bunch spacing ({\unit{ns}}) & 100--200 & 25 & 75--150 \\
    Average pileup & 0.25 & 3 & 0.0004 \\
    Peak instantaneous luminosity ({\percms}) & $8 \ten{29}$ & $1 \ten{34}$ & $6 \ten{27}$ \\
    Integrated luminosity ({\nbinv}) & $1.73 \ten{2}$ & $3.04 \ten{5}$ & 1.70 \\
    Peak interaction rate ({\unit{kHz}}) & 2\ten{3} & 4\ten{4} & 50 \\
 \end{tabular}
}
\label{tab:collrate}
\end{table}

\begin{figure*}[htb!]
\centering
\includegraphics[width=\cmsFigWidthVeryWide]{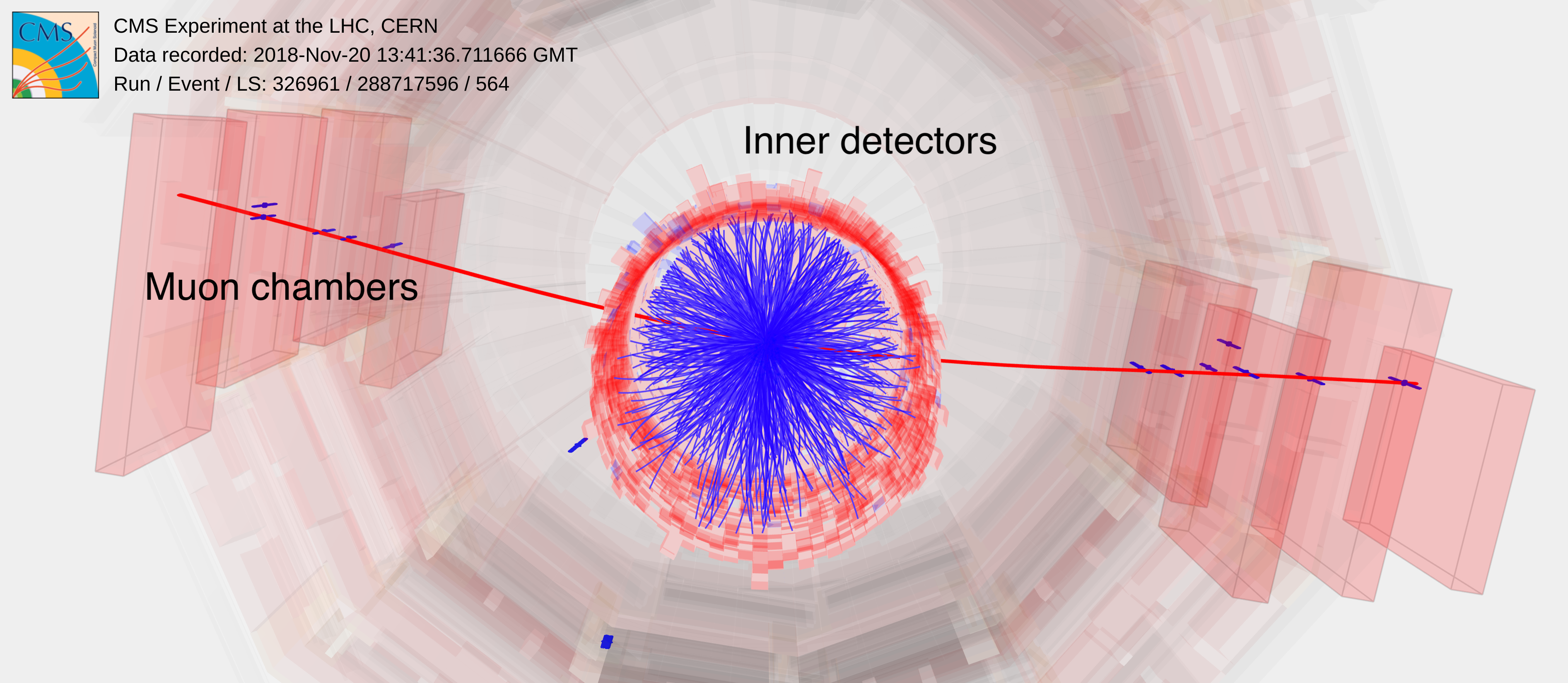}
\caption{Dimuon event from a \PbPb collision in the CMS detector. Particles reconstructed in the inner tracker are colored blue; two muons from a \Z boson candidate are colored red. See text for details.}
\label{fig:eventDisplay}
\end{figure*}

A typical dimuon event in the heavy ion environment is displayed in Fig.~\ref{fig:eventDisplay}. The inner part of the detector reconstructs the large number of tracks produced in the collision, shown as blue lines in the figure. The electromagnetic and hadronic energy captured by the calorimeters is shown as the semitransparent towers surrounding the inner tracks (the endcap calorimeters are hidden for clarity). Most of the event activity is contained within the inner detectors. Two muons are displayed as two long red tracks reaching the outermost part of the detector, where the muon chambers are located. The high occupancy in the inner detector might impact the performance of the pixel detector and silicon strip tracker and reduce the muon tracking efficiency. 

Most tracks in \PbPb collisions originate from a single primary vertex, in contrast to high-luminosity \pp collisions, where the tracks originate from vertices with different positions along the beam line. The separation between vertices in \pp collisions is used to mitigate the effect of pileup on the reconstruction. The single-vertex origin, together with higher multiplicity, are factors that were addressed in the reconstruction of \PbPb events.

\subsection{Muon triggers in the heavy ion environment}
\label{sec:trigger}

An important component of all muon measurements involves an efficient selection of collision events that contain muons. The real-time (online) muon reconstruction algorithms in heavy ion collisions are generally identical to those used in \pp collisions. This is true for both the L1 trigger and the HLT. There are three different types of muons defined in the CMS trigger system, commonly referred to as L1, L2, and L3 muons. An L1 muon is identified by the hardware-based L1 trigger system using solely the muon detectors. The L2 and L3 muons are defined within the HLT: an L2 muon uses only information from the muon detectors, while an L3 muon additionally matches the track from the muon detectors to a track from the inner tracker. In each subsequent level, the momentum resolution and purity (defined in Section~\ref{sec:ZTrig}) are improved, as studied in Ref.~\cite{MUO-19-001}.

The heavy ion runs are characterized by a much smaller instantaneous luminosity and interaction rate when compared with 13\TeV \pp runs. The lower rate allows the use of looser trigger requirements in \PbPb: lower \pt thresholds are used for muons, as well as lower quality requirements (\eg, no isolation criterion). Relaxing the requirements benefits the physics program that is based on heavy ion muon data, focusing on open heavy flavor, quarkonium, and electroweak boson production; particularly by increasing the quarkonium acceptance in the low-\pt part of the spectrum.

Two main triggers used in analyses involving muons are studied in this paper: a high-\pt single-muon trigger (designed for the measurements of electroweak bosons and top quarks), and a low-\pt double-muon trigger (for quarkonia and \PB mesons). Each of these triggers is described in more detail in Sections~\ref{sec:ZTrig} (high-\pt single-muon) and \ref{sec:JpsiTrig} (low-\pt double-muon). To quantify the performance of these triggers, we collected an independent sample of muons using monitoring triggers. The monitoring triggers were single-muon HLT triggers with varying \pt thresholds. The single-muon triggers were used because any muon that satisfies the trigger can be used as a tag, and any other muon in the event can be used as a probe, since it has no trigger requirement imposed on it. Thus, the monitoring triggers allowed us to collect a muon-rich sample that is minimally biased for the purpose of the tag-and-probe analyses.

\section{Extraction of efficiencies}
\label{sec:extraction}

We express the total muon efficiency, $\epsilon_\mu$, as the product of three contributions: muon reconstruction efficiency (\ereco), identification efficiency (\eID), and triggering efficiency (\etrig), such that:
\begin{equation}
\epsilon_\mu = \ereco\,\eID\,\etrig
\label{eq:factorization}
\end{equation}
Each efficiency is calculated as the fraction of muons that pass the specific selection in Eq.~\eqref{eq:factorization}. The numerator in one step (\eg, muon passing the identification requirements) is used as a denominator in the next step (\eg, trigger), ensuring the factorization of the total efficiency. Therefore, the particular efficiencies are relative rather than absolute, showing the efficiency with respect to the previous step. In the case of the first step, \ereco, the denominator consists of tracks from the silicon tracker with minimal quality requirements, such that very few genuine muons are removed from the denominator. The efficiency related to finding the tracks in the inner tracker, \etrack, is investigated separately (see Ref.~\cite{CMS:TRK11-001}, where \etrack is obtained for \pp collisions using simulated single muons). For purposes of this paper it is not included in Eq.~\eqref{eq:factorization}. The variable \ereco quantifies the reconstruction efficiency in the outer muon detectors and the efficiency of matching tracks from the silicon tracker to tracks from the muon detectors.

\begin{figure*}[htb!]
\centering
\includegraphics[width=\cmsFigWidthVeryWide]{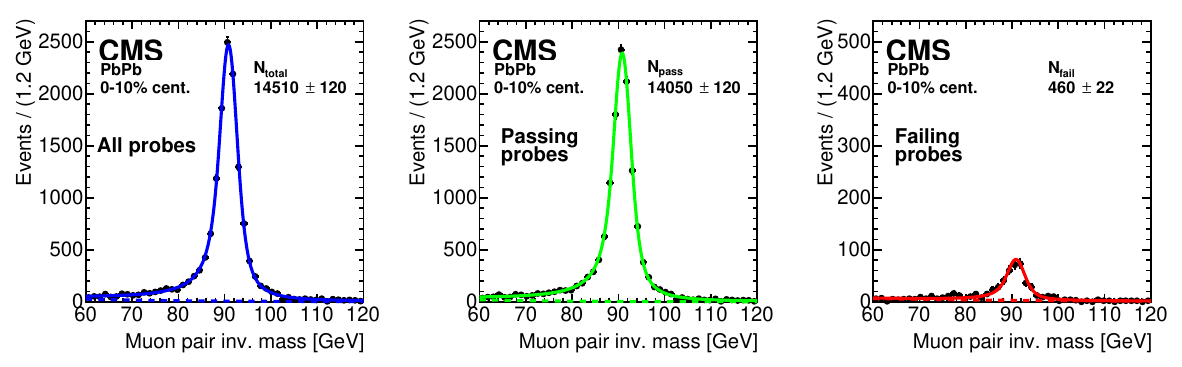}
\caption{Example of a fit to the data in \PbPb collisions. The three panels show invariant mass distributions of the tag-and-probe pairs in central 0--10\% collisions fitted with the sum of signal and background components. Left panel: total spectrum. Middle panel: spectrum for pairs where the probe passed the muon identification selection. Right panel: spectrum for muon pairs where the probe failed the selection. The vertical scale of the failing probes is enhanced by a factor of 5 compared with the other panels for visibility. The yields displayed in the legends of the figures correspond to the \Z boson signal as extracted from the fitting procedure described in the text. The efficiency is retrieved directly from the simultaneous fit and in this particular case is $\epsilon = 0.968 \pm 0.002$. }
\label{fig:exampleFit}
\end{figure*}

The muon efficiencies are calculated using the data-driven tag-and-probe method (an example of its use in CMS is presented in Ref.~\cite{ppMuon_2012}). First, we select events where a resonance, such as the \JPsi meson or the \Z boson, was likely to have decayed into a muon pair, producing a distinct peak in the dimuon invariant mass distribution. In these events, the `tag' muon is a muon that satisfies strict requirements, such that its probability of being a genuine muon is as high as possible. Once we have `tagged' a muon with a high level of confidence, we pair it with a second muon picked with much more relaxed requirements, in order to `probe' its properties. The requirements placed on the probe muon define the denominator of the efficiency and are thus chosen in accordance with the factorization in Eq.~\eqref{eq:factorization}. With one tag muon and one probe muon, we construct tag-probe pairs and obtain an invariant mass distribution. The left panel of Fig.~\ref{fig:exampleFit} shows an example of this invariant mass distribution, in which we have reconstructed the \Z boson mass peak to study the probe muons in the kinematic region of $15 < \pt <200 \GeV$. Next, we apply to the probes the selection criteria that we want to study. The subset of muons that pass these selection criteria are labeled `passing probes', which then define the numerator in the efficiency. Those muons that fail the selection in question are dubbed `failing probes'. We construct invariant mass distributions for each, seen in the middle panel of Fig.~\ref{fig:exampleFit} for passing probes, and in the right panel for failing probes. 

We perform a simultaneous fit to passing and failing distributions to measure the efficiency. The simultaneous fit includes both a signal and a background component for each invariant mass distribution. Only contributions from the signal peaks are considered in the efficiency calculation. The signal yields from the fit are denoted as $N_{\text{total}}$, $N_{\text{pass}}$, and $N_{\text{fail}}$ for total, passing, and failing probes, respectively. Since the sum of passing and failing probes equals the total of all probe muons contained in the signal peak, the corresponding yields are related as follows: 
\begin{align}
N_{\text{pass}} &= \epsilon \, N_{\text{total}} 
\label{eq:passing}\\
N_{\text{fail}} &= (1 - \epsilon) \, N_{\text{total}} 
\label{eq:failing}
\end{align}
We can obtain the efficiency, $\epsilon$, by appropriately including it as a parameter in the simultaneous fit. The statistical uncertainty of the efficiency is calculated directly in the fitting procedure, and takes into account the uncertainties in the parameters that describe both the signal and the background. The systematic uncertainty is obtained by choosing alternative functions to describe the signal and background shapes and by varying the fitting range. The systematic uncertainty obtained from the signal-shape variation is typically ${<}0.1\%$ when binning the data as a function of \Ntracks. The corresponding uncertainty for the background-shape variation is ${<}0.4\%$, and for the fitting-range variation it is ${<}0.4\%$. When studying the uncertainty in a specific kinematic region, the uncertainty in a few cases can reach up to a few percent. The comparisons among the different collision systems are not significantly affected by the systematic uncertainties, given their small size. Furthermore, we analyzed the fits such as those in Fig.~\ref{fig:exampleFit} across all collision systems using the same signal shape and fitting ranges to minimize the effects of the signal and mass range selection on the efficiency comparisons. The remaining source of systematic uncertainty, the background shape, can be different due to the varying amounts of combinatorial background for each collision system. Overall, the effects of systematic uncertainty are smaller than the size of the statistical uncertainty in the data, and are therefore not shown in the figures. For the MC simulations, the sizes of the statistical uncertainties are generally smaller than the symbols.

Events with muon pairs from decays of \Z boson and \JPsi resonances are used to measure the efficiencies. Because of the different mass of these particles, the decay muons have different \pt spectra, allowing for the study of efficiencies in different kinematic regimes. In particular, muons from the \Z boson peak are used in the \pt range 15--200\GeV, whereas the muons from the \JPsi peak are used for studies in the low-\pt range 0.8--25\GeV. The efficiency corrections obtained from the tag-and-probe method are then applied to each muon that enters a CMS analysis, under the assumption that the muon detection efficiency is independent of the type of parent particle that produces the decay muon. This assumption holds when the two decay muons are sufficiently separated in $(\eta,\phi)$, where $\phi$ is the azimuthal angle in radians, such that the detection of the one muon is independent of the detection of the other. The efficiencies can also be used for muons that come from a secondary vertex (such as those from nonprompt \JPsi decays) as long as the analysis applies selections that are consistent with those used to derive the efficiencies. With these considerations, the tag-and-probe method is used in Section~\ref{sec:trackingEff} for studying the muon reconstruction efficiency, and in Sections~\ref{sec:Zpeak} and \ref{sec:Jpsi} for identification and trigger efficiencies using the \Z and \JPsi resonances, respectively.

\section{Muon reconstruction}
\label{sec:trackingEff}

The algorithms for offline muon reconstruction for 13\TeV \pp data in Run 2 are documented in Ref.~\cite{MUO-16-001}. The same algorithms are used for the reconstruction of \pPb data, as well as of \pp reference data for heavy ions. The algorithms used in 2018 \PbPb data taking were based on the \pp algorithms with only minor customizations related primarily to the track selection. The customized \pp algorithms worked in the \PbPb environment, despite the large increase in detector occupancy, because the addition of a fourth layer to the pixel detector, as mentioned in Section \ref{subsec:alignment}. The fourth layer allowed the use of higher quality four-hit seeds for the first tracking step, which significantly reduced the rate of spurious combinations of hits and thus the processing time needed per event.

The CMS muon reconstruction starts by creating `tracker tracks', which are tracks reconstructed using only information from the inner tracker; and `standalone muons', which are reconstructed only from the muon system. Two approaches are then used \cite{ CosmicMuon, ppMuon_2012}: 
\begin{itemize}
	\item \textit{Global muon reconstruction (outside-in)}: Reconstruction starts with a standalone muon, which is 	then used to look for possible track matches in the inner tracker. A `global muon' is then obtained from fitting the combined hits of the two track components.
	\item \textit{Tracker muon reconstruction (inside-out)}: A complementary approach starts with all tracker tracks with $\pt > 0.5\GeV$ ($\pt > 0.8\GeV$ for \PbPb collisions) and momentum $p > 2.5\GeV$. These are extrapolated to the muon system, and if at least one matching muon segment is found, the track is qualified as a `tracker muon'. The principal advantage of this approach is that it reconstructs lower-momentum muons more efficiently than the outside-in method. The drawback is that the algorithm starts with a much higher number of tracks, as the occupancy in the tracker is much larger than in the muon detectors.
\end{itemize}
Almost all of the muons (${\approx}99\%$) in the detector's kinematic range are successfully reconstructed as global or tracker muons~\cite{ppMuon_2012}. 

\begin{figure}[htb!]
\centering
\includegraphics[width=\cmsFigWidthWide]{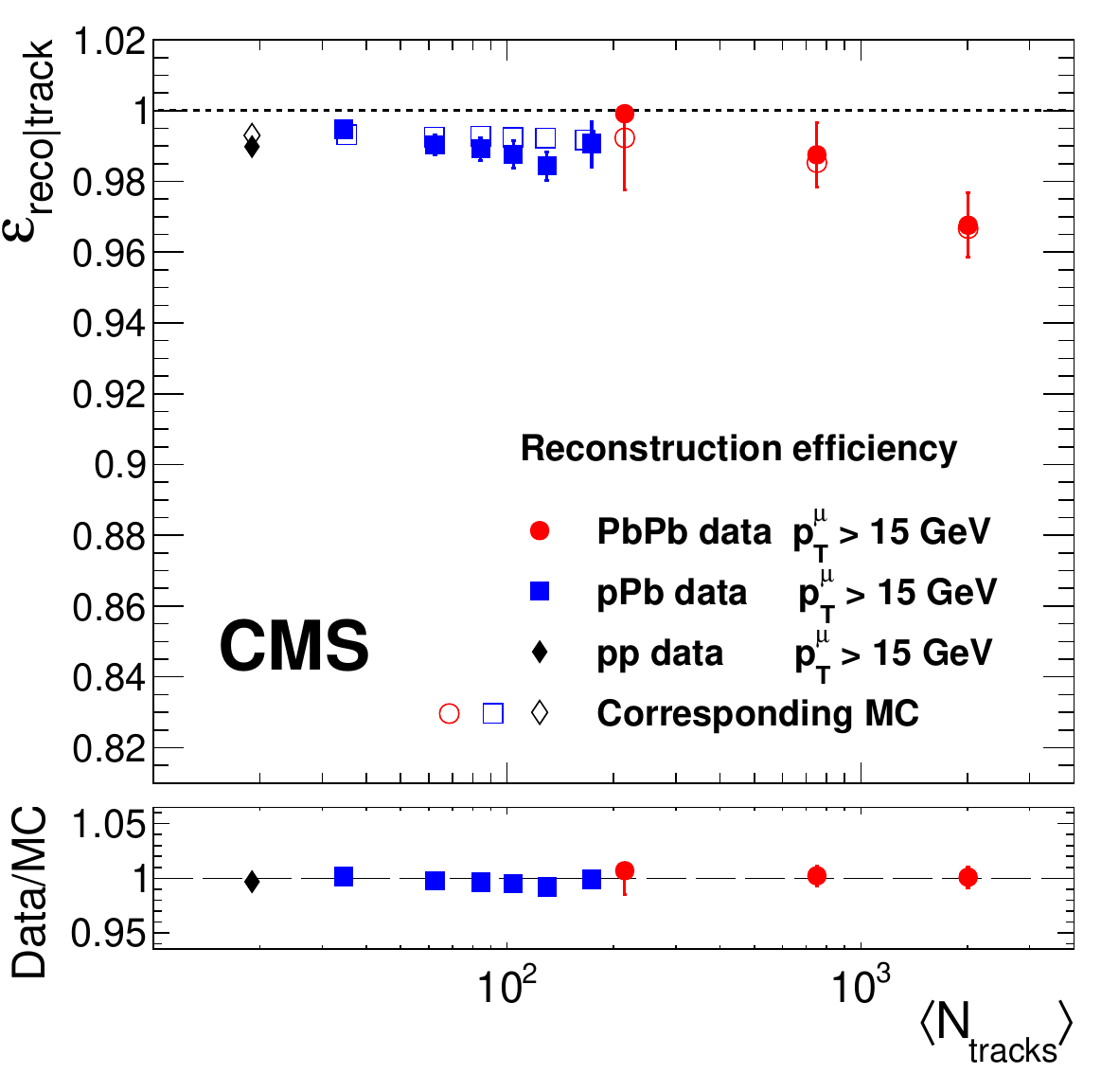}
\caption{Muon reconstruction efficiency (defined as the probability that a given muon that produced a tracker track will be reconstructed as both a global and a PF muon) plotted as a function of the number of tracks in \pp, \pPb, and \PbPb collisions. Open symbols are the MC results corresponding to each data set. Lower panel shows the ratio between data and MC simulation. Only statistical uncertainties are shown.}
\label{fig:MuonReconstruction}
\end{figure}

Reconstructed muons are used as input into the CMS particle-flow (PF) algorithm~\cite{CMS_PFalgo}, which combines information from all CMS subdetectors to identify and reconstruct all individual particles for each event, including electrons, photons, neutral hadrons, charged hadrons, and muons. For muons, the PF algorithm applies a set of selection criteria to candidates reconstructed with the standalone, global, or tracker muon algorithms~\cite{MUO-16-001}.

Figure~\ref{fig:MuonReconstruction} shows the reconstruction efficiency \ereco for muons with $\pt>15 \GeV$ in \pp, \pPb, and \PbPb collisions as a function of \Ntracks, obtained from fits to the \Z boson resonance. In this case, probes are defined as any track reconstructed in the inner tracking system with a set of basic quality requirements (at least one hit in the pixel detector and a signal in at least six tracker layers). The passing probe satisfies both global-muon and PF-muon requirements. The reconstruction has high efficiency (${\gtrsim}97\%$) across the range of occupancies, although it exhibits a minor deterioration towards the most central \PbPb collisions, both in collision data (filled points) and MC simulations (open points). We note that the track multiplicity increases by about 2 orders of magnitude from \pp to central \PbPb collisions, while the efficiency drops only 2 percentage points. To obtain the tracking efficiency in the inner tracker specifically, which can be added to Eq.~\eqref{eq:factorization} to account for the tracking efficiency factor, \etrack, we can use a standalone muon as a probe and a corresponding inner track as a passing probe. From this complementary approach (not shown) we find that the inner tracking efficiency is high (${\gtrsim}97\%$) even in the large multiplicity environment. The results indicate that the CMS muon reconstruction is robust against large increases in the inner detector occupancy.

\section{Studies in the \texorpdfstring{\Z}{Z} boson kinematic region}
\label{sec:Zpeak}

The dimuon decay of the \Z resonance is used to study the efficiencies for muons in the kinematic range $15 < \pt < 200\GeV$. The \pt range is bounded by the kinematics of the \Z boson decay muons and by the number of such decays in the studied data set. The efficiencies obtained from the \Z boson peak region have been used for several high-momentum analyses in \pPb and \PbPb collisions, such as the studies of \Z boson~\cite{Z_2021, CMS:2021otx}, \PW boson~\cite{W_2012}, and \ttbar production via $\cPqt \to  \PWp\cPqb \to \Pgmp\cPgn\,\cPqb$~\cite{TopHIN-19-001}. Table~\ref{tab:procedureBoth} summarizes the settings that are commonly used in CMS analyses with muons in the \Z boson kinematic region, and in the \JPsi kinematic region (Section~\ref{sec:Jpsi}). Following Eq.~\eqref{eq:factorization}, muon identification efficiency in the \Z boson kinematic range starts with the set of muons that satisfy both the global and the PF muon requirement (\eID denominator). This is done because the set constitutes the passing probes in the previous step (\ereco numerator). Section~\ref{sec:ZID} explains the muon ID and Section~\ref{sec:ZTrig} discusses the trigger. 

\begin{table*}[htb]
\renewcommand{\arraystretch}{1.0} 
  \centering
	\topcaption{Overview of muon identification and triggers used in the \Z boson kinematic region (\eg, \Z, \PW, and \cPqt analyses) and in the \JPsi kinematic region (\eg, \JPsi and \PgU analyses). A description of muon identification and triggers is given in the text.}
	\cmsTable{
  \begin{tabular}{l c  c c}
   Region & Collision system &  Muon ID        &  Trigger  \\ \hline
	\Z &\pp/\pPb/\PbPb        		& Tight        &  Single muon; $\pt > 12 \GeV$; decision at HLT      \\ [\cmsTabSkip]
	\multirow{3}{*}{\JPsi}    &\pp               & Hybrid-soft       & Double muon; no \pt requirement; decision at L1 \\ 
     &\pPb              & Soft             & Double muon; no \pt requirement; decision at L1   \\ 
     &\PbPb         		& Hybrid-soft       & Double muon; \pt, mass, and $\Delta R$ requirements; decision at HLT     \\ 
  \end{tabular}
	}
	\label{tab:procedureBoth}
\end{table*}

\subsection{Muon identification efficiency}
\label{sec:ZID}

The majority of the CMS heavy ion results that are using muons in the 15--200\GeV \pt range apply the `tight-muon' selection, which is identical to that used in \pp analyses~\cite{ppMuon_2012}. The selection is as follows:
\begin{itemize}
\item Global muon and PF muon: The muon is reconstructed as both a global and a PF muon, as outlined in Section~\ref{sec:trackingEff}. 
\item Fit $\chi^2/\mathrm{dof} < 10$ and number of muon hits ${>}0$: The global muon fit must be of good quality to reduce the number of muons from nonprompt hadron decays and the number of hadrons that traverse the inner detectors and pass through the magnet, reaching the muon chambers (hadronic punch-through). The selection places a limit on the fit $\chi^2/\mathrm{dof}$ (where dof stands for degrees of freedom) and requires at least one muon detector hit included in the fit.
\item Number of matched muon stations ${\geq}2$: At least two muon stations must be matched to the muon to further reduce hadronic punch-through.
\item Number of pixel hits ${>}0$ and number of strip layers ${>}5$: To further improve the muon track quality and provide a good \pt measurement, minimal numbers of pixel hits and strip layers with hits are required.
\item $\Dxy < 0.2\unit{cm}$ and $\Dz < 0.5\unit{cm}$: An upper limit on the distance of closest approach between the primary vertex and the muon track in the transverse plane, \Dxy, is used. This requirement selects muons from prompt decays and reduces contributions from cosmic rays and from nonprompt decays of hadrons. The corresponding limit in the longitudinal direction \Dz is larger, due to coarser resolution of the  tracker in the z~coordinate.
\end{itemize}

\begin{figure}[htb]
\centering
\includegraphics[width=\cmsFigWidthWide]{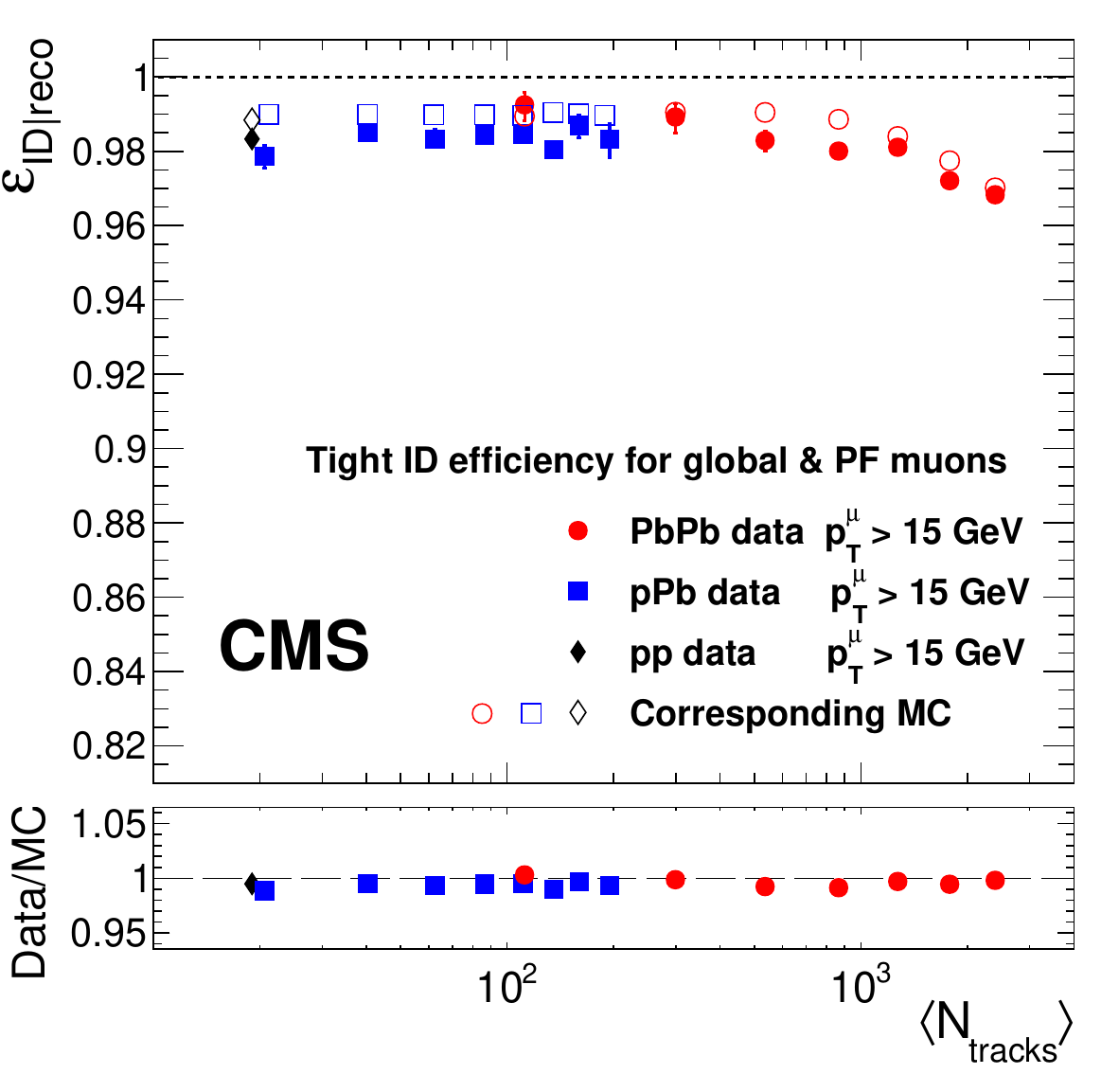}
\caption{Tight ID efficiency for global and PF muons as a function of the number of tracks in \pp, \pPb, and \PbPb collisions. Open symbols are the MC results corresponding to each data set. Lower panel shows the ratio between data and MC simulation. Only statistical uncertainties are shown.}
\label{fig:TightIDComp2}
\end{figure}

\begin{figure}[htb]
\centering
\includegraphics[width=\cmsFigWidth]{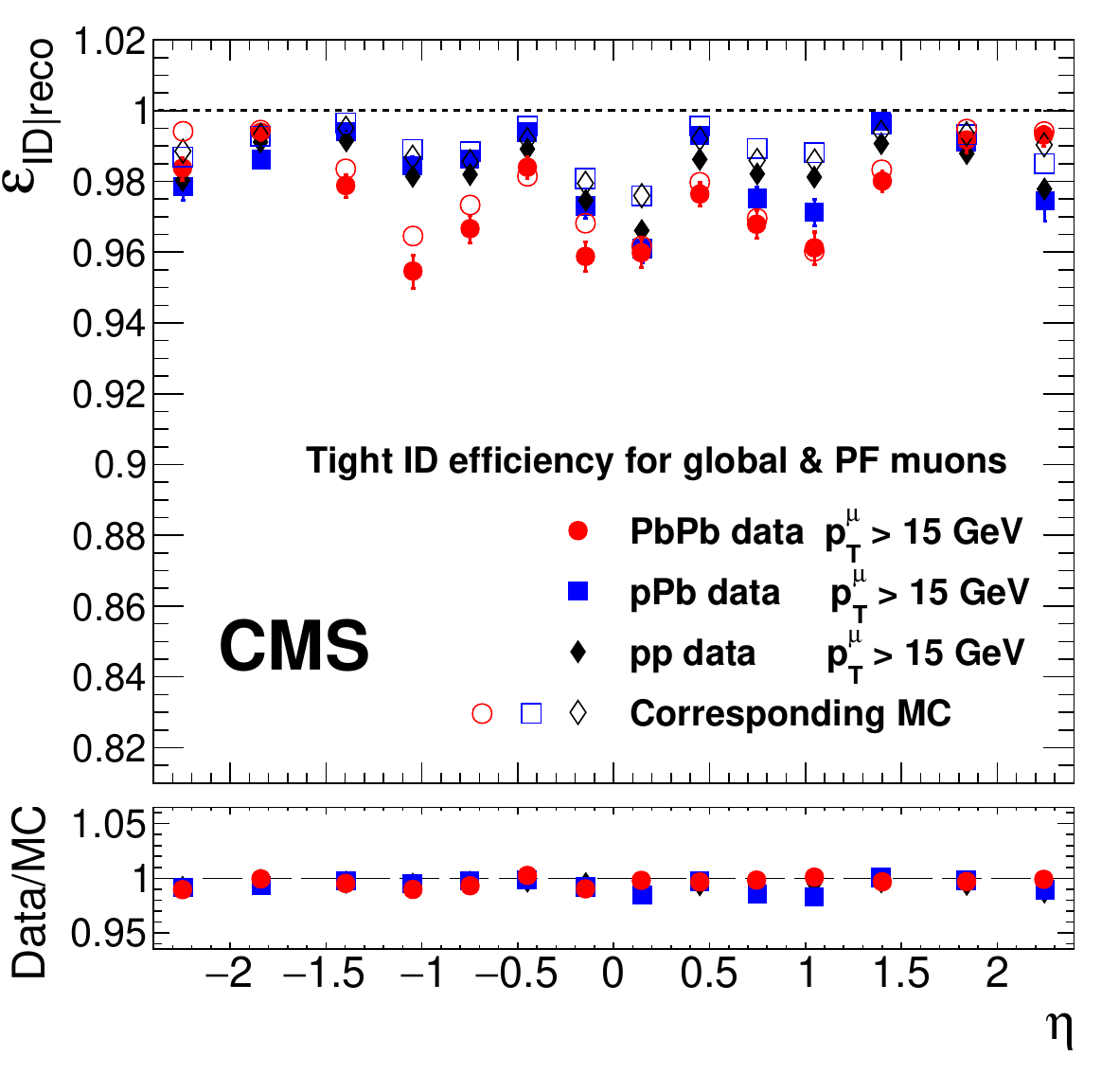}
\includegraphics[width=\cmsFigWidth]{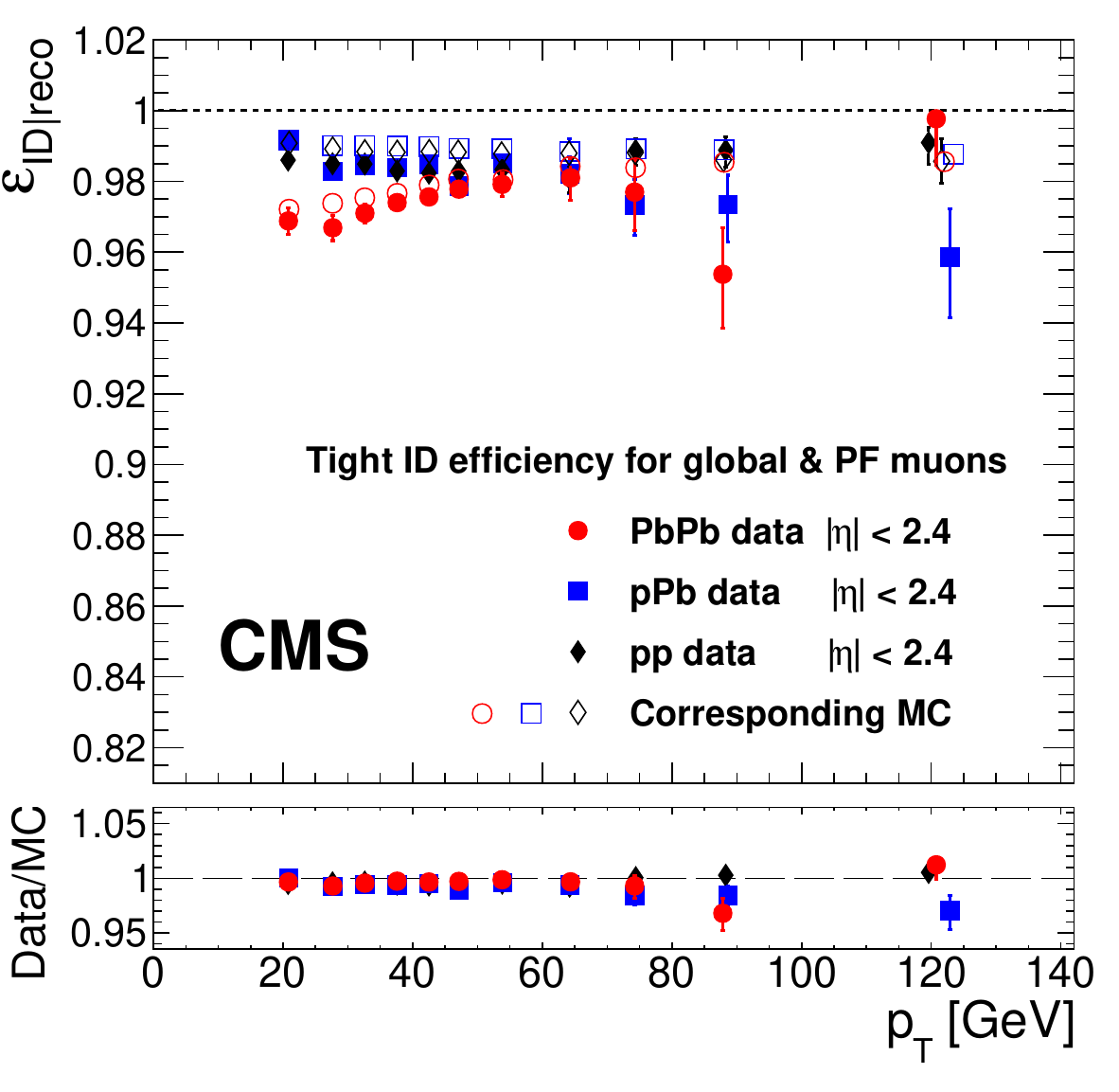}
\caption{Tight ID efficiency as a function of $\eta$ (\cmsLeft) and \pt (\cmsRight) in \pp, \pPb, and \PbPb collisions. The probe is both a global muon and a PF muon. Open symbols are the MC results corresponding to each data set. Lower panels show the ratio between data and MC simulation. Only statistical uncertainties are shown.}
\label{fig:TightIDComp1}
\end{figure}

Figure \ref{fig:TightIDComp2} displays the \Ntracks dependence of the tight muon ID efficiency. The efficiencies in \pp, \pPb, and \PbPb collisions are in good agreement with each other. The values are consistent with a constant in the full range of \Ntracks reached in \pp and \pPb collisions, but exhibit a minor drop for the central \PbPb collisions. Even with track multiplicity over $2000$, the efficiency only falls to about 96\%. Figure~\ref{fig:TightIDComp1} shows the kinematic dependence. The efficiency in \PbPb collisions is comparable with that from lower multiplicity collisions in most of the studied kinematic range. However, the \PbPb efficiency is 1--2\% lower in the region of $15 < \pt \lesssim 40\GeV$ and in the barrel-endcap overlap region ($ 0.9  < \abs{\eta} < 1.2$). This loss of efficiency is caused primarily by the tight ID requirement of at least two matched muon stations. This requirement is met by fewer muons when the occupancy is high, which is the reason for the slight efficiency decrease in central collisions seen in Fig.~\ref{fig:TightIDComp2}. The two-matched-stations requirement is also the primary cause for the small difference between the data and MC results in both Figs.~\ref{fig:TightIDComp2} and \ref{fig:TightIDComp1}. Even though the MC description of the tight muon identification slightly overestimates its efficiency, it captures all the trends visible in data.

\subsection{Muon trigger efficiency and purity}
\label{sec:ZTrig}

The tag-and-probe method is used to measure the efficiency of both trigger levels (L1+HLT) combined. In all the collision systems, the HLT studied here is a single-muon trigger with a requirement of $\pt > 12 \GeV$ at the HLT. The L1 seed was tailored to the specific conditions of each data-taking run. In \PbPb collisions, the seed has a \pt threshold of $3 \GeV$; in \pp and \pPb collisions only L1 muons with $\pt > 7 \GeV$ were considered by the HLT.

\begin{figure}[htb]
\centering
\includegraphics[width=\cmsFigWidthWide]{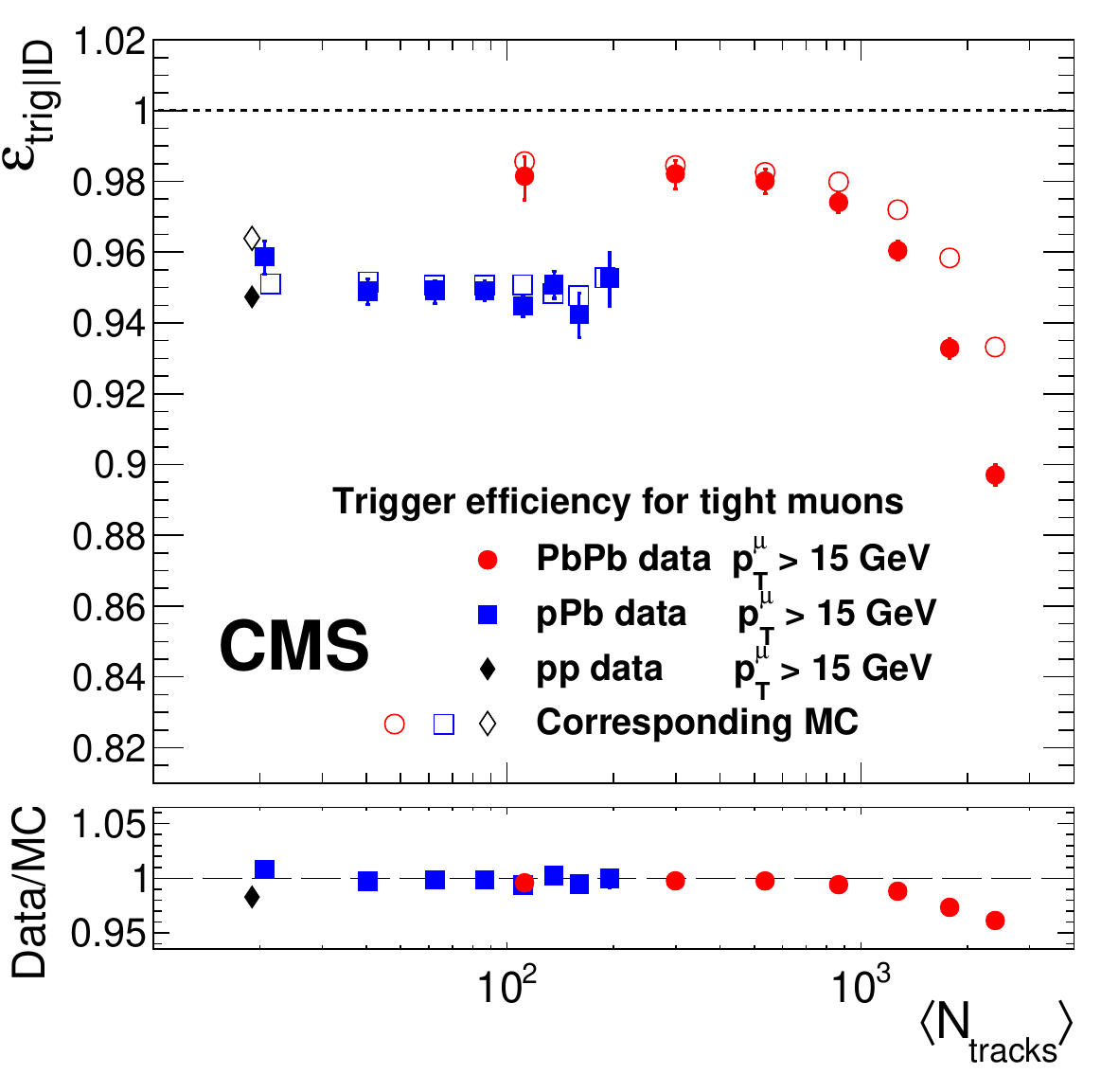}
\caption{Trigger efficiency of tight muons as a function of the number of tracks. The trigger requires a single muon with \pt above $12\GeV$. The efficiency is calculated for muons with $\pt>15\GeV$ to avoid threshold effects. Open symbols are the MC results corresponding to each data set. Lower panel shows the ratio between data and MC simulation. Only statistical uncertainties are shown. The significant increase in the trigger efficiency for the PbPb data set occurs because of detector upgrades and trigger optimizations prior to the 2018 data-taking period.}
\label{fig:TriggerComp2}
\end{figure}

\begin{figure}[htb]
\centering
\includegraphics[width=\cmsFigWidth]{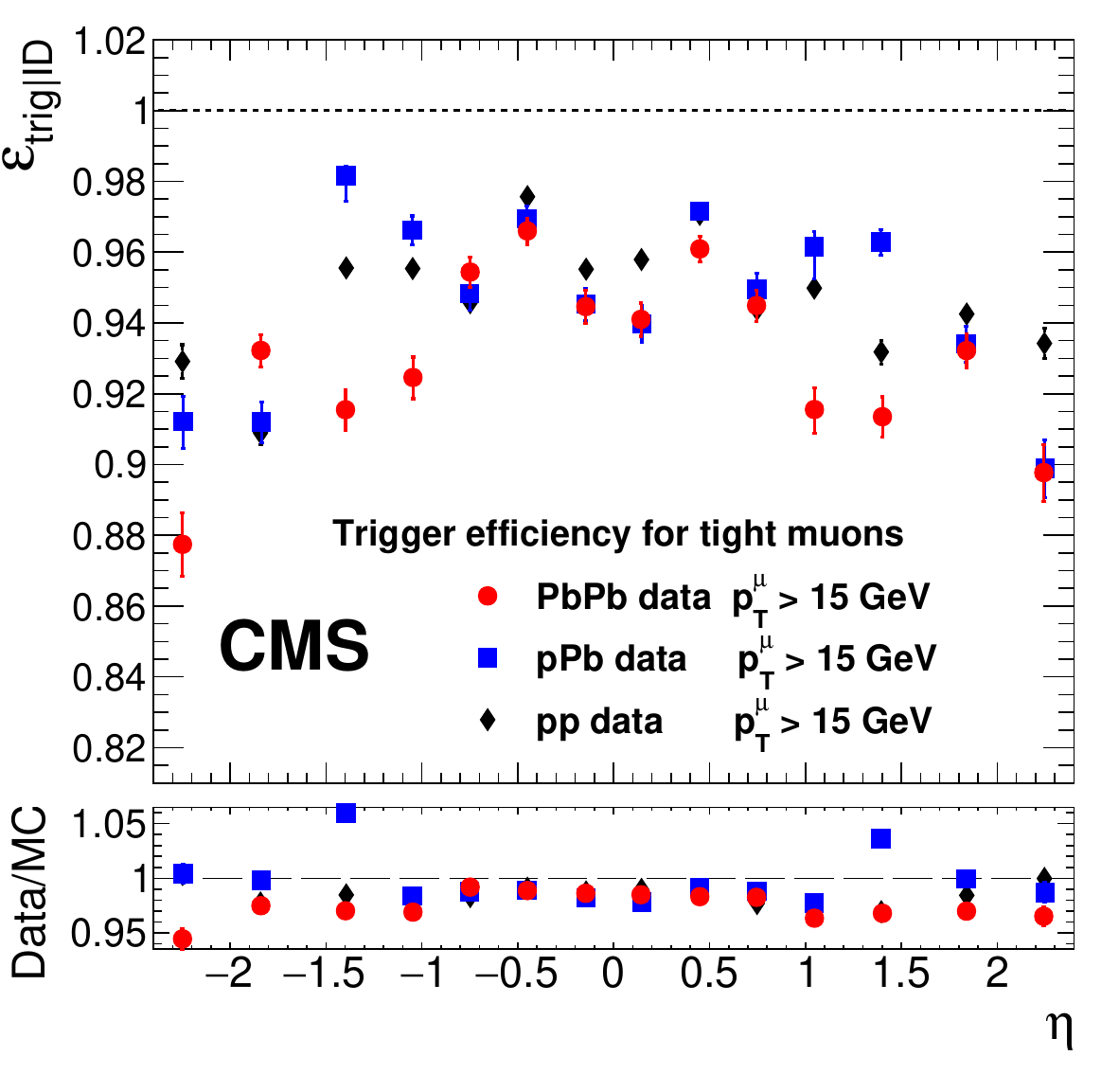}
\includegraphics[width=\cmsFigWidth]{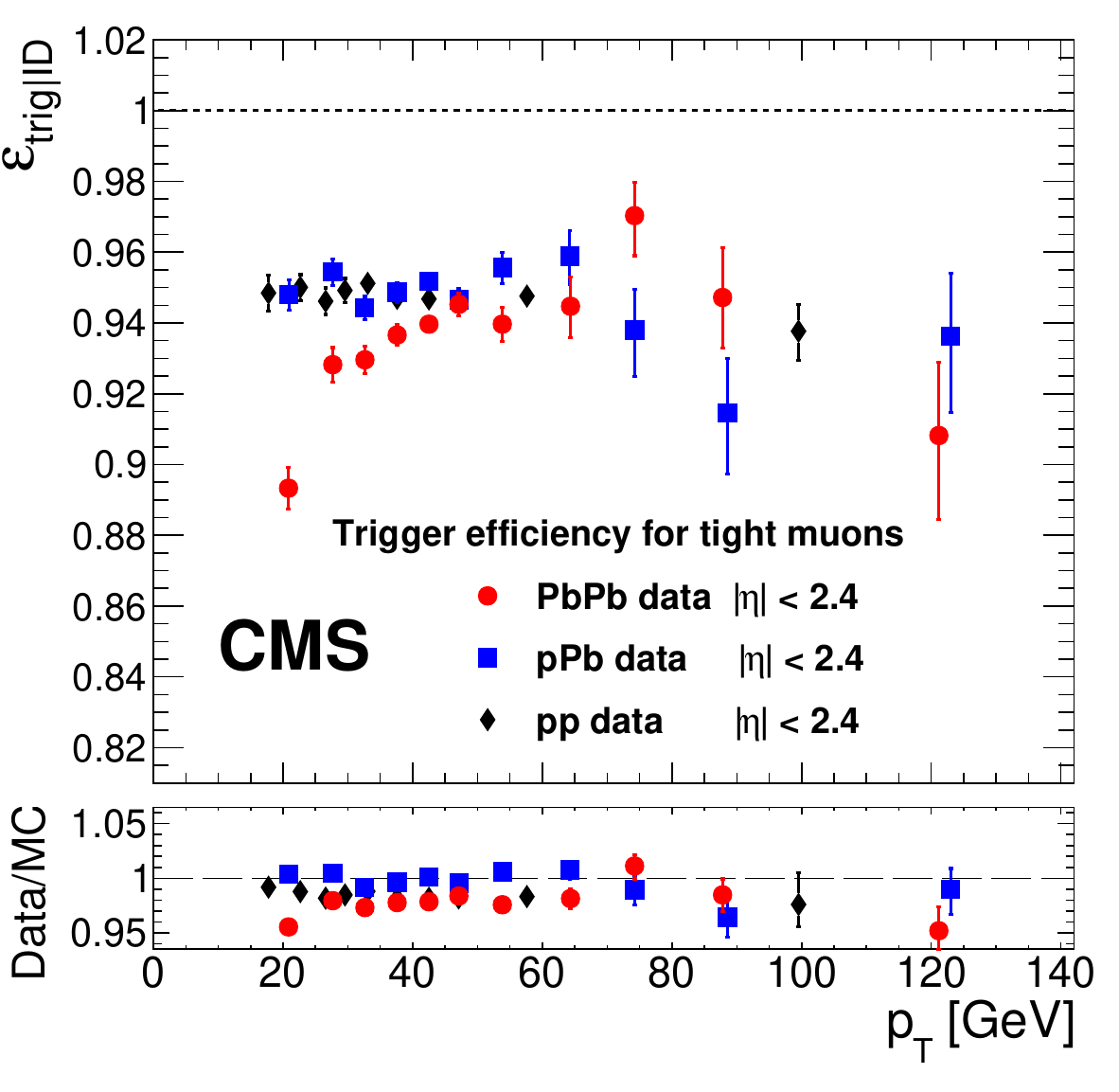}
\caption{Trigger efficiency of tight muons as a function of $\eta$ (\cmsLeft) and \pt (\cmsRight) in \pp, \pPb, and \PbPb collisions. Lower panels show the ratio between data and MC simulation (MC points are omitted in the upper panels for clarity). Only statistical uncertainties are shown.}
\label{fig:TriggerComp1}
\end{figure}

Figure \ref{fig:TriggerComp2} shows the trigger efficiency as a function of \Ntracks. The probes for the trigger efficiency (denominator) are the tight muons; the passing probes (numerator) are those tight muons that also pass the trigger requirement, following the factorization scheme of Eq.~\eqref{eq:factorization}. The quantitative comparison among all three data sets is affected by the upgrades to the detector between each corresponding run. In addition, the 2018 \PbPb data set is affected by the optimizations of the HLT that were done between 2017 and 2018~\cite{MUO-19-001}. The combination of the detector upgrades and trigger optimizations increased the overall efficiency in \PbPb data when compared with the \pp and \pPb data taken in previous years, as seen by the efficiency shift from \pPb to peripheral \PbPb in the figure. With these changes taken into consideration, the efficiency is effectively flat in the \Ntracks range covered by \pp and \pPb collisions, but begins to decline in more central \PbPb events after the number of tracks in the detector surpasses several hundred. This drop is more pronounced for the trigger efficiency than was observed for muon ID (Fig.~\ref{fig:TightIDComp2}), suggesting that the former is more sensitive to detector occupancy than the latter. The MC simulations (open symbols) describe the general trends in data fairly well, however the trigger simulation for \PbPb collisions overestimates the efficiency for central events. Figure~\ref{fig:TriggerComp1} shows the efficiency dependence on the muon kinematics. We observe similar trends to those for the ID efficiency: the trigger in \PbPb collisions is less efficient in the intermediate pseudorapidity region ($1.0<\abs{\eta}<1.6$). While the efficiency stays mostly flat as a function of \pt in \pp and \pPb events, it decreases towards low \pt in \PbPb events. In both \pp and \pPb collisions the trigger efficiency reaches a plateau at ${\approx}95\%$ for muons with $\pt>20\GeV$. The \PbPb muons at 20\GeV have a trigger efficiency of about 89\%; the efficiency rises until $\pt \approx 40\GeV$, at which point it reaches a plateau. From the data to MC simulation ratio, we can infer that the simulation captures the trends fairly well, but overestimates the efficiency in \PbPb events for muons with $\pt\lesssim 70 \GeV$. The $\eta$-dependence of the data/simulation ratio for \PbPb collisions is reasonably flat; the overall shift in the results of the simulation is caused by the overestimation in the high-\Ntracks low-\pt region that was previously noted. 

\begin{figure}[htb]
\centering
\includegraphics[width=\cmsFigWidthWide]{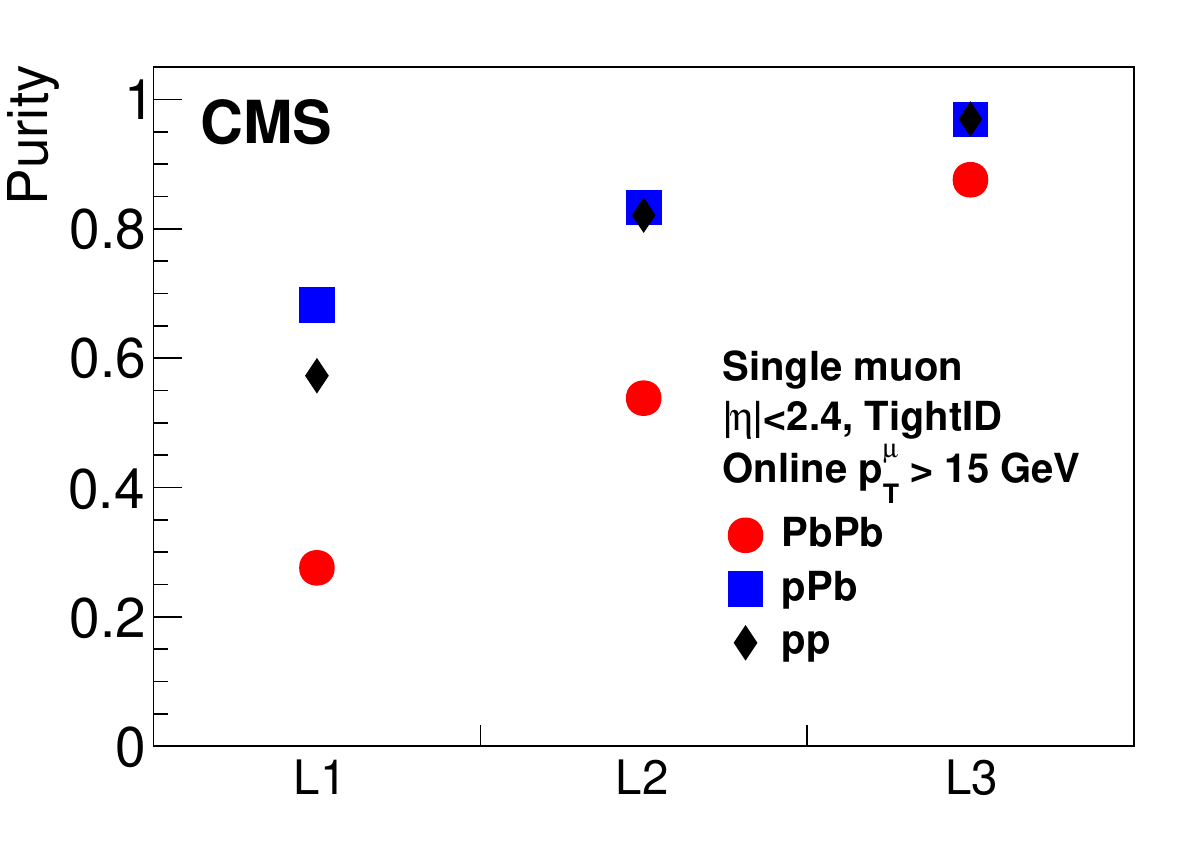}
\caption{Purity for the L1, L2, and L3 trigger steps, compared among \pp, \pPb, and \PbPb collisions. The online muon must have $\pt > 15\GeV$ and the offline muon matched to it must pass the tight ID selection and have $\abs{\eta}<2.4$. Details of the purity definition are given in the text. Statistical uncertainties are smaller than the symbol sizes.}
\label{fig:purity}
\end{figure}

The trigger purity, which quantifies the capability to avoid triggering on particles that are not genuine prompt muons, is an important component for the determination of the trigger rate. We explore the purity of the three online muon definitions described in Section~\ref{sec:trigger}---L1, L2, and L3---to illustrate improvements in purity at each subsequent trigger stage. We define the purity as a ratio of the numbers of muons in the two samples. The sample in the denominator is the set of online muons passing the relevant trigger step and having an online $\pt>15\GeV$ requirement in order to focus on muons from the \Z boson decay. The numerator is meant to be genuine muons, which in our case are the subset of online muons consisting of those matched geometrically within $\Delta R \equiv\sqrt{\smash[b]{(\Delta\phi)^2 + (\Delta\eta)^2}}<0.3$ (0.1 for L3) to an offline muon that must satisfy the tight ID requirement. Offline muons must have $\abs{\eta}<2.4$ without an additional \pt requirement. For our study, muons were collected with dimuon triggers that utilize L1/L2/L3 trigger selections. Although the exact values of the purity are dependent on the details of the trigger that was used to collect the muon sample, the study provides useful qualitative comparisons among the collision systems and trigger steps. The results of the purity study are shown in Fig.~\ref{fig:purity}. The purity values in \pp and \pPb collisions are similar, whereas the purity in \PbPb collisions is lower. The figure also illustrates how the purity improves in each successive and more complex trigger step, thus reducing the rate. Although the rate is reduced, the HLT maintains a very high efficiency, as seen earlier in Fig.~\ref{fig:TriggerComp2}. Backgrounds to the purity include primarily muons from long-lived hadron decays. For the L3 trigger, the background also includes muons where a track from the muon chambers is matched to one inner track online, but offline it is matched to a different track. This occurs more often in high multiplicity \PbPb collisions.

\section{Studies in the \texorpdfstring{J/$\boldsymbol{\psi}$}{J/psi} kinematic region}
\label{sec:Jpsi}

\begin{figure}[htb]
\centering
\includegraphics[width=\cmsFigWidth]{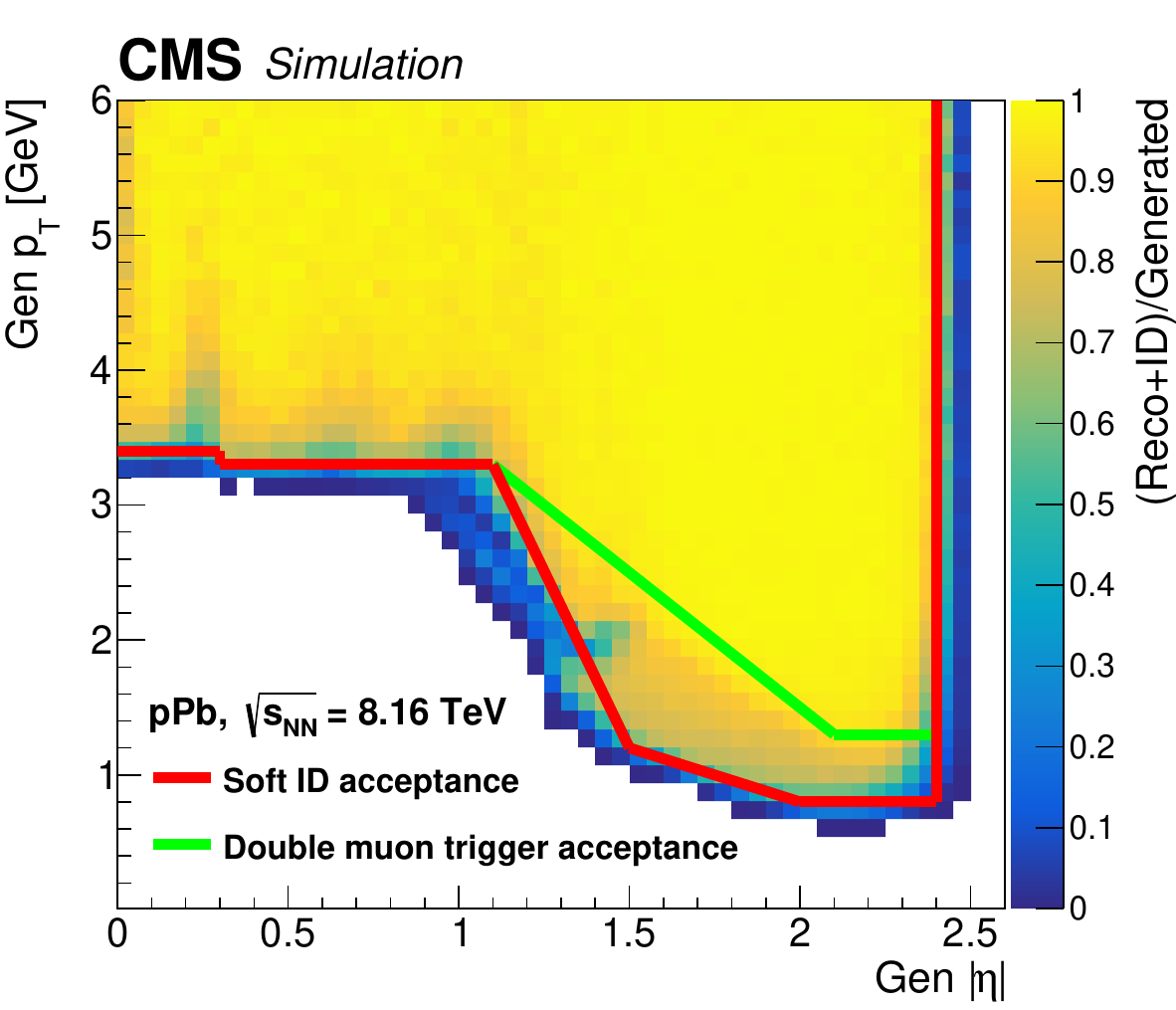}
\includegraphics[width=\cmsFigWidth]{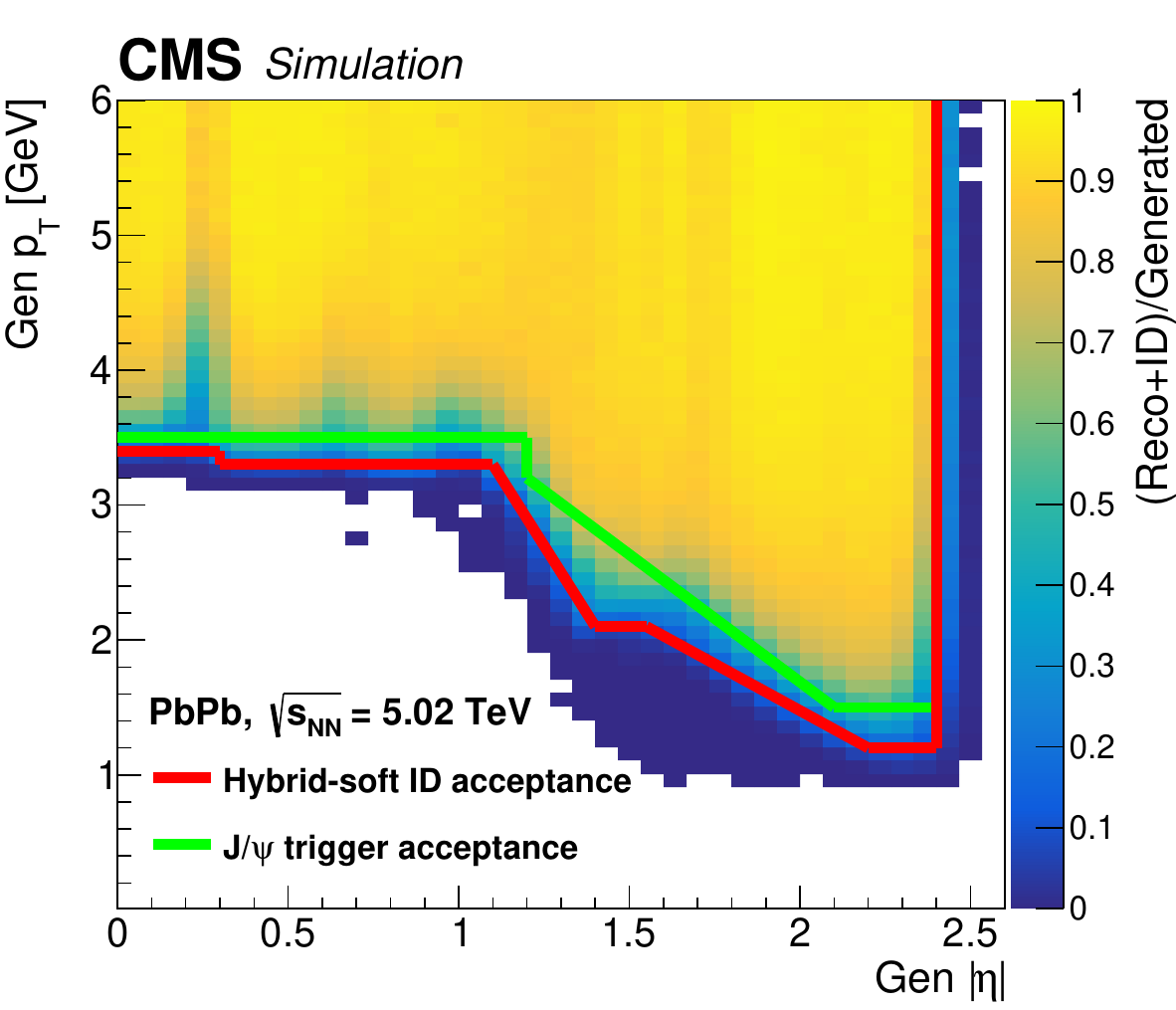}
\caption{Regions of the CMS detector commonly used in the heavy ion muon analyses for \pPb (\cmsLeft) and \PbPb (\cmsRight) collisions. For each panel, the combined reconstruction and identification efficiency for simulated muons is plotted as a function of generated muon $\abs{\eta}$ and \pt. The lower-threshold curves (red) are for muon identification, and are used only by those analyses that do not use a dedicated muon trigger. The higher-threshold curves (green) are used by most analyses (those using the muon trigger information).}
\label{fig:Acceptance}
\end{figure}

The \JPsi resonance is used to derive the efficiencies for low-\pt muons. The acceptance at low-\pt is limited by the material in front of the muon stations~\cite{CMS-TDR} and by the strength of the CMS magnetic field. Figure \ref{fig:Acceptance} details the acceptance regions commonly used for the various collision systems. The green curves, with higher-\pt thresholds, pertain to analyses that use the most frequently employed muon trigger in heavy ion collisions. The red curves, with lower thresholds in \pt, are used for muon selection by those analyses that use only muon identification and do not use a dedicated muon trigger (\eg, they use a jet trigger or a minimum-bias trigger to collect their data samples). The existence of the two distinct acceptance regions in each collision system arises from the better low-\pt efficiency of muon identification compared with that of the muon trigger. The lower portion of Table~\ref{tab:procedureBoth} summarizes muon ID selections and triggers commonly used in the \JPsi kinematic region in the various collision systems. The muon ID selections are explored in Section~\ref{sec:JpsiID} and the triggers are discussed in Section~\ref{sec:JpsiTrig}.

\subsection{Muon identification efficiency}
\label{sec:JpsiID}

The `soft muon' ID \cite{ppMuon_2012, MUO-16-001} denotes a set of selection criteria that is commonly used within CMS for low-\pt muons. The selection has been further optimized for the heavy ion environment and the resulting selection is called `hybrid-soft muon'. The hybrid-soft muon ID is used in \PbPb analyses, as well as for \pp reference data, and consists of the following requirements:
\begin{itemize}
\item Tracker muon and global muon: The muon is reconstructed as both a tracker and a global muon (defined in Section~\ref{sec:trackingEff}) to reduce the misidentification rate.
\item Number of pixel layers ${>}0$ and number of strip layers with hits ${>}5$: To further improve muon-track quality, we impose a minimum on the numbers of pixel layers and strip layers associated with the track.
\item $\Dxy < 0.3\unit{cm}$ and $\Dz < 20\unit{cm}$: An upper limit on the distance of closest approach between the event vertex and the muon track in the transverse plane, \Dxy, selects muons from prompt decays, and reduces contributions from cosmic rays and from nonprompt hadron decays. The variable \Dz is the corresponding longitudinal component of the distance of closest approach, and the requirement on \Dz is very loose.
\end{itemize}
The soft muon ID differs from the hybrid-soft ID by having a high-purity requirement~\cite{Chatrchyan:2014fea} applied to the tracker tracks, instead of requiring the muon to be a global muon. The soft muon ID is used in \pPb and other \pp collisions in general, as it allows reaching lower-\pt than does hybrid-soft ID. The soft muon ID is not ideal for \PbPb collision data, since the high-purity requirement has a low efficiency in the central collisions.

\begin{figure}[htb]
\centering
\includegraphics[width=\cmsFigWidth]{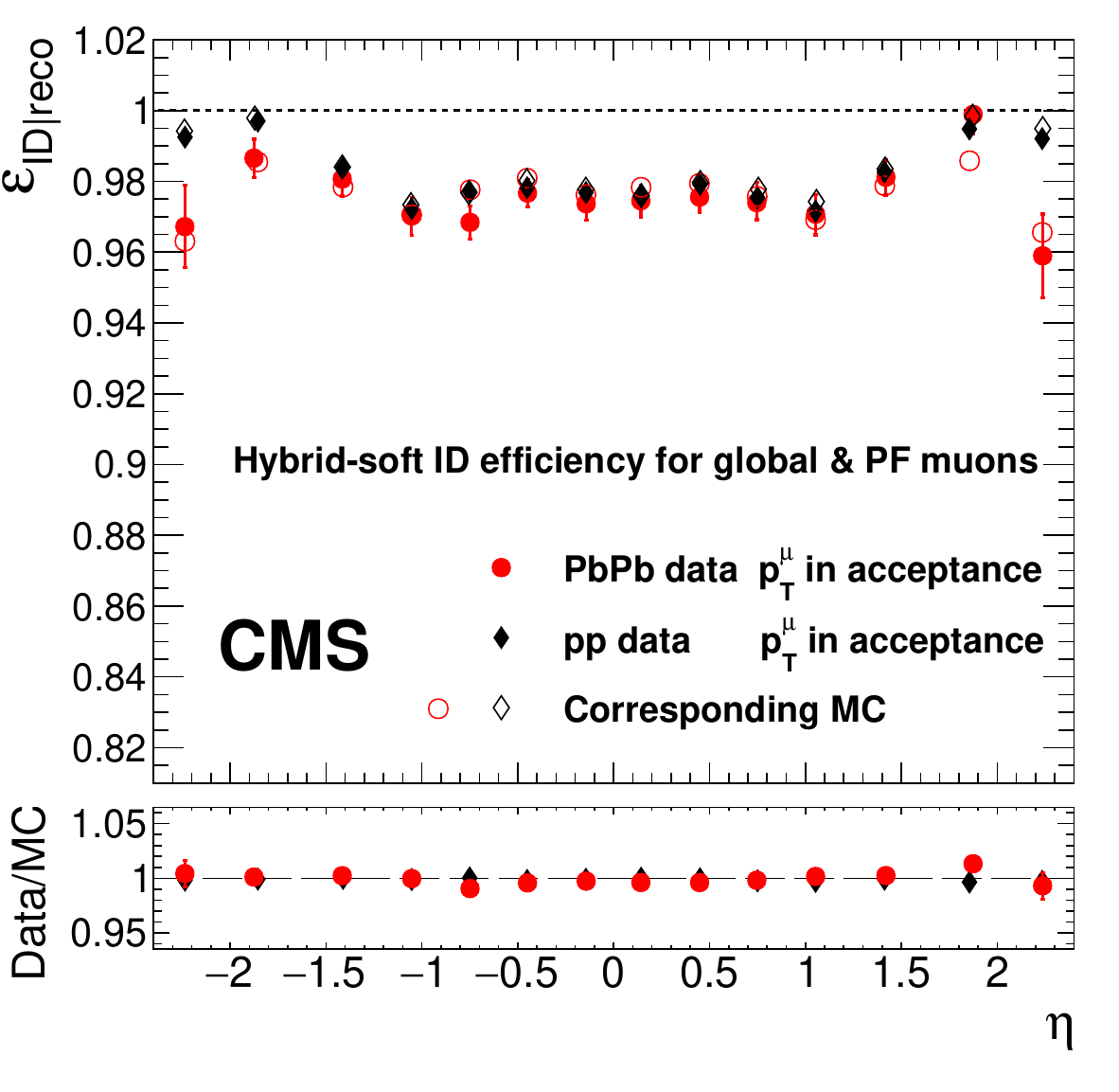}
\includegraphics[width=\cmsFigWidth]{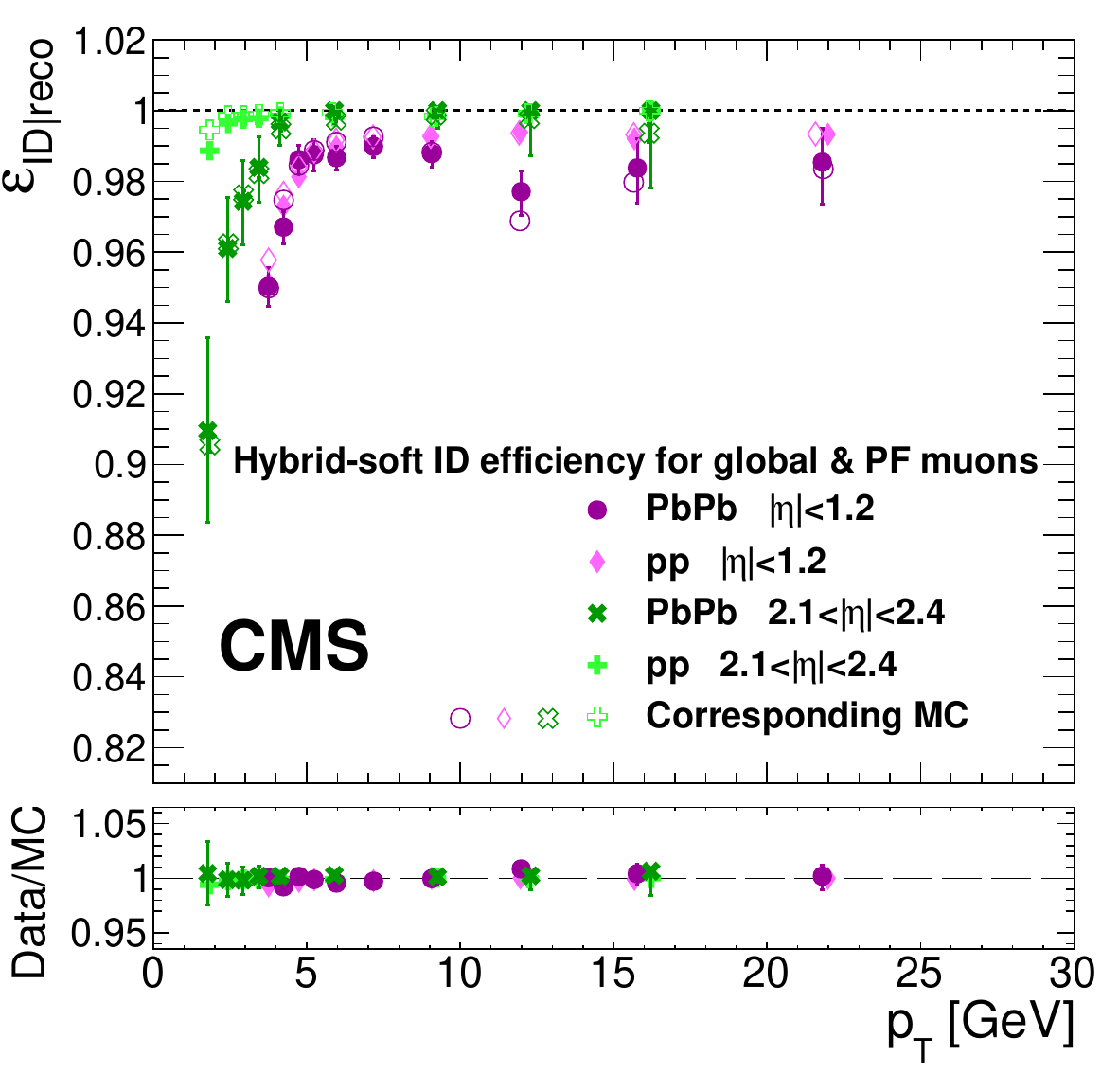}
\caption{Hybrid-soft ID efficiency for global muons as a function of $\eta$ (\cmsLeft) and \pt (\cmsRight) in \pp and \PbPb collisions. The muons are restricted in acceptance as shown by the red line in the \cmsRight panel of Fig.~\ref{fig:Acceptance}. Lower panels show the ratio between data and MC simulation. Only statistical uncertainties are shown.}
\label{fig:SoftIDComp1}
\end{figure}

Figure \ref{fig:SoftIDComp1} shows the pseudorapidity and \pt dependence of the efficiency of the hybrid-soft selection. The kinematic reach at low \pt is better in the endcap region (as seen in Fig.~\ref{fig:Acceptance}). The efficiency is comparable between the \pp and \PbPb collisions, except for the region of $\abs{\eta}>2$ and $\pt < 5\GeV$, where the occupancy in the tracker is the highest. We found that the drop occurs primarily in central \PbPb events (not shown), in which the high number of tracks in the tracker leads to a decrease in matching efficiency with the standalone muons. This trend is well-described by the MC simulations (open points), indicating that they successfully model this behavior. 

\subsection{Muon triggers}
\label{sec:JpsiTrig}

The muon trigger used for quarkonia (\eg,~\JPsi and \PgU mesons) studies in \pPb and for reference \pp data sets is as loose as possible: a double-muon trigger at L1. It simply required two tracks in the muon system, with only very loose quality requirements and no explicit \pt threshold. The trigger decision is taken at L1 with no further filtering at the HLT stage. In \PbPb collisions, the rate of this double-muon trigger at L1 was above the acceptable limit, so that additional quality requirements had to be placed at the HLT level to reduce the rate. One of the two muons was required to be an L2 muon; the other was required to be an L3 muon. This was done as a compromise between reducing the rate and keeping the efficiency as high as possible for low-\pt quarkonia. Two main HLT paths were used: one for charmonia and one for bottomonia. The charmonia path required the invariant mass of the dimuon to be within 1 and 5\GeV, and required a $\Delta R<3.5$ between the two muons to suppress uncorrelated muons. The bottomonia path required a minimum mass of 7\GeV to cover the bottomonia mass region around 9.4--10.5\GeV, and minimum \pt requirements of 2.0\GeV for the L2 muon and 2.5\GeV for the L3 muon. Both triggers required that the muons in the pair have opposite charge. To complement the two main HLT trigger paths, centrality-gated triggers were used to select peripheral events, which have the smallest occupancy in \PbPb collisions. Therefore, the data-stream size was able to be kept sufficiently low to satisfy bandwidth and storage limitations. The differences among the triggers used in \pp, \pPb, and \PbPb collisions preclude meaningful comparisons of the trigger efficiency.

\section{Mass resolution and scale}
\label{sec:scaleAndRes}

The masses of the \Z boson and the \JPsi meson are well known from other results~\cite{Zyla:2020zbs}. Measuring these particles via their dimuon decay can provide additional information about the muon reconstruction. The dimuon mass resolution provides a suitable window into the precision of the momentum measurement; the mean value of the mass peak is sensitive to the momentum scale of the detector. The momentum calibration procedure for our data sets is described in Ref.~\cite{CMS:TRK-20-001}. The mass resolution is obtained from the width of the fitted dimuon resonance. We apply the same selection as we did for the trigger-efficiency evaluation, described in Sections~\ref{sec:Zpeak} and \ref{sec:Jpsi} for each corresponding resonance. To study the absolute scale of the momentum calibration, we plot the reconstructed mass of the resonance obtained from the fit, \mfit, scaled by the world-average mass obtained by the Particle Data Group (PDG), \mPDG~\cite{Zyla:2020zbs}. Since measurements of particle masses using heavy ion data are rare, a small bias in the momentum scale has had little effect on the success of the overall program. In contrast, heavy-ion measurements are more sensitive to momentum resolution. For example, a degraded resolution would impact the signal-to-background ratio for resonances.

\begin{figure}[htb!]
\centering
\includegraphics[width=\cmsFigWidth]{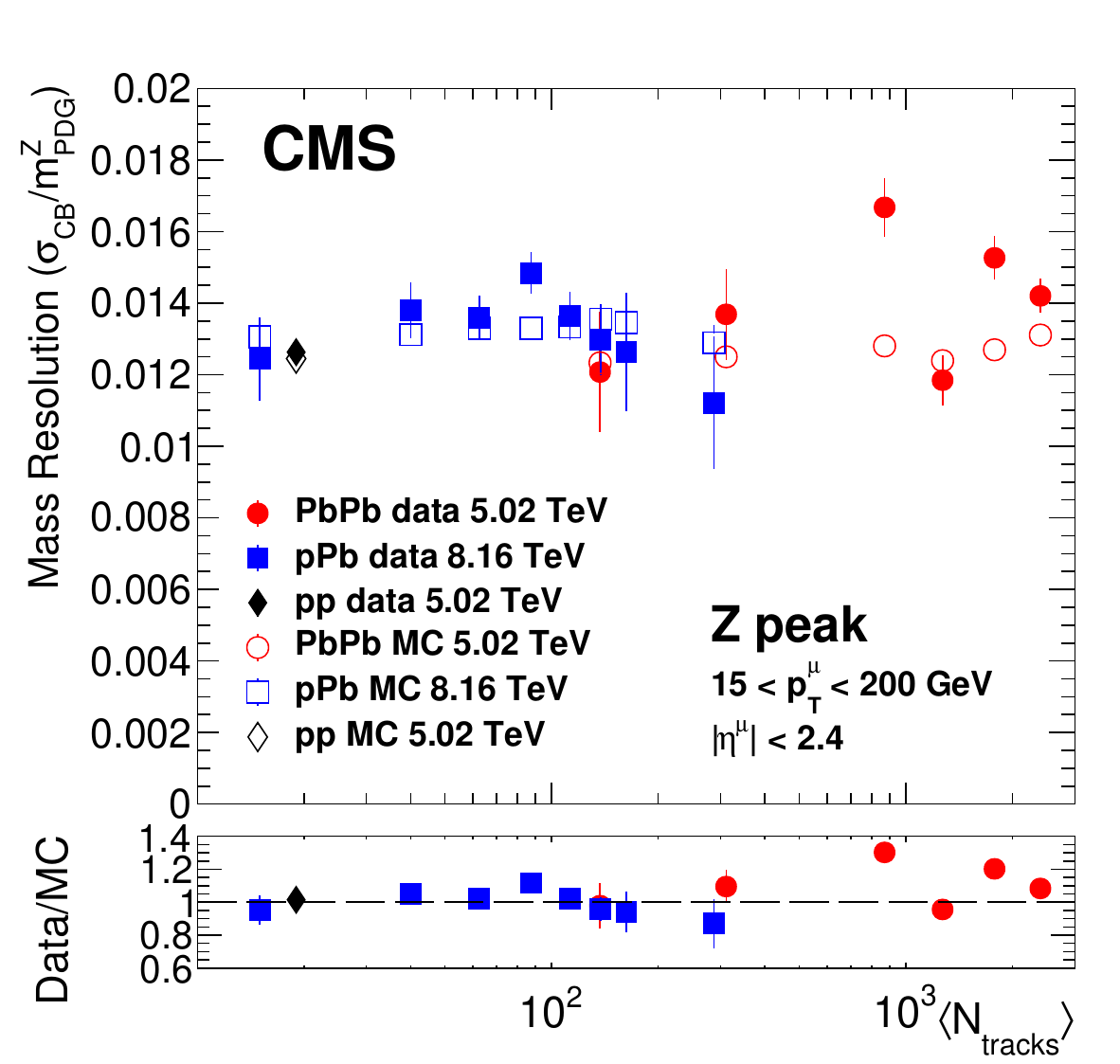}
\includegraphics[width=\cmsFigWidth]{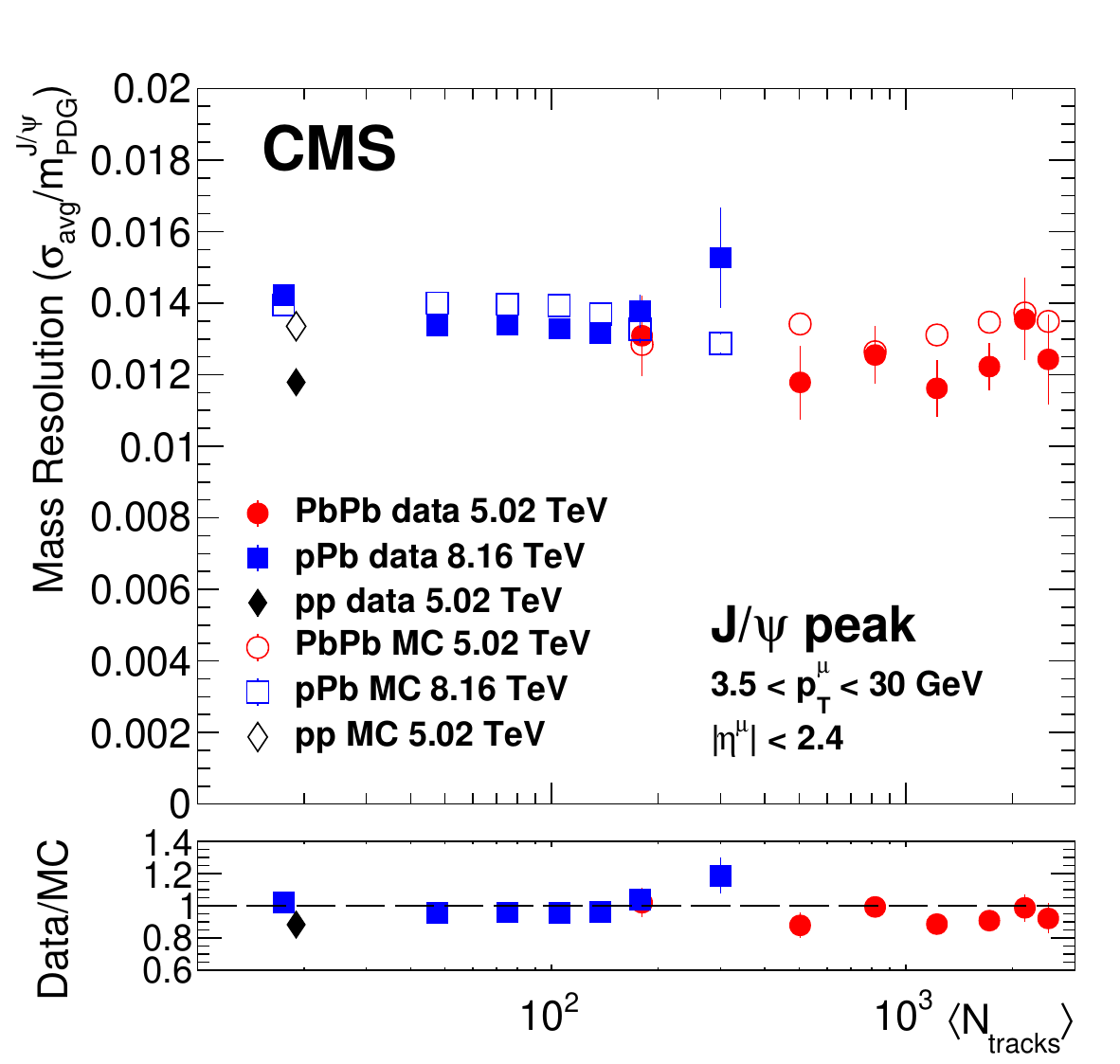}
   \caption{Mass resolution at the \Z boson peak (\cmsLeft) and the \JPsi peak (\cmsRight) as a function of the number of tracks from MC simulations (open points) and collision data (filled points) in \pp, \pPb, and \PbPb collisions. The mass resolution is scaled by the world-average mass \mPDG. Lower panels show the ratio between data and MC simulation. Only statistical uncertainties are shown.}
\label{fig:MassResolution}
\end{figure}

For the \Z boson, the signal peak is described by a Breit--Wigner (BW) function convolved with a Crystal Ball (CB) function. The CB function consists of a Gaussian peak with a power-law tail grafted onto the lower-valued side, such that the function and its derivative are continuous~\cite{SLAC-R-236}. In this case, the reconstructed mass, \mfit, is a common parameter between the CB and the BW functions that is both the mean of the Gaussian core of the CB function and the location parameter of the BW function. The width of the BW function is fixed to the natural width of the \Z boson ($\Gamma_Z\approx2.5\GeV$). The Gaussian width of the CB function, \sigmaCB, is a free parameter in the fit and summarizes our information about the muon resolution in the detector. We plot the parameter \sigmaCB divided by the \Z boson PDG mass, \mPDG. 

\begin{figure}[htb!]
\centering
\includegraphics[width=\cmsFigWidth]{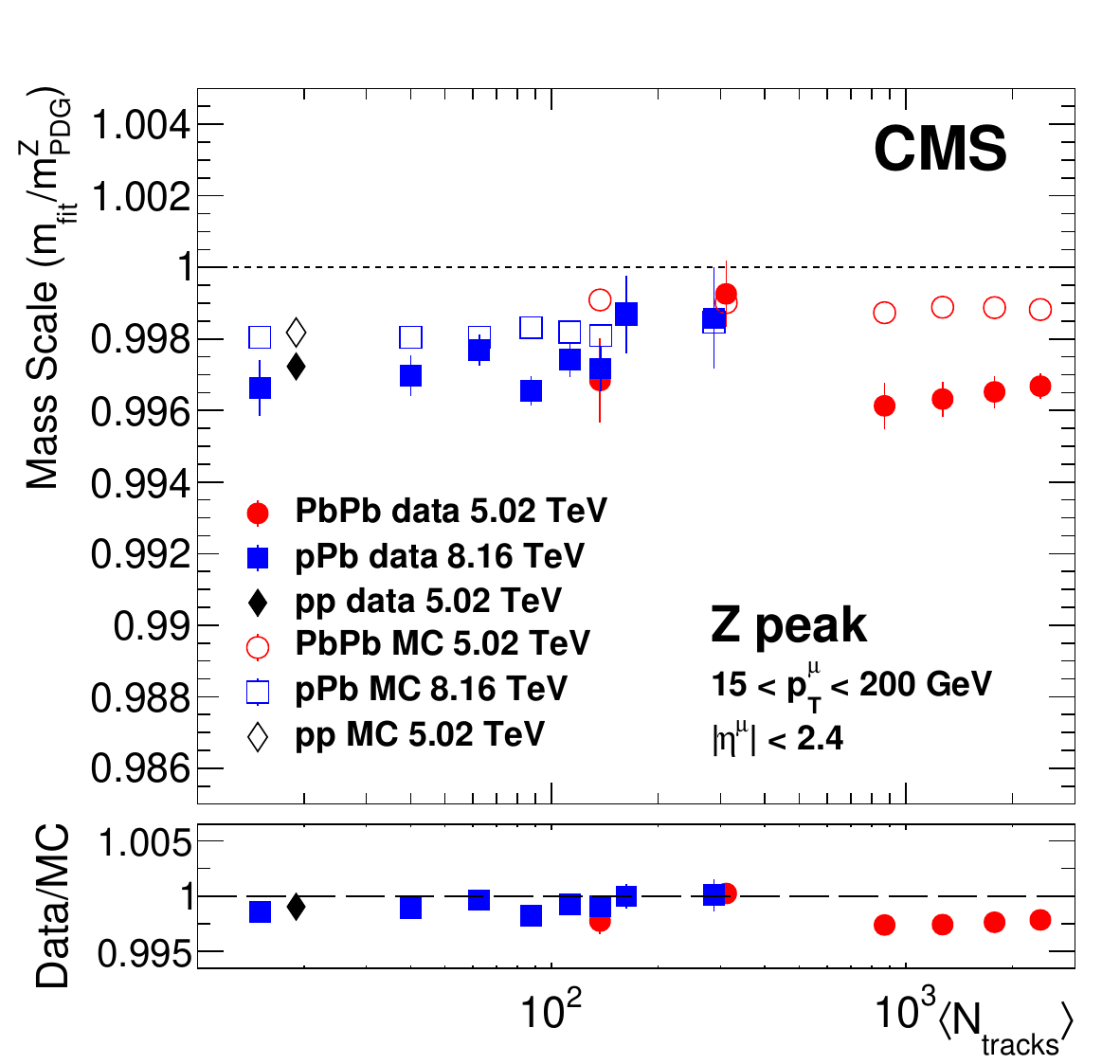}
\includegraphics[width=\cmsFigWidth]{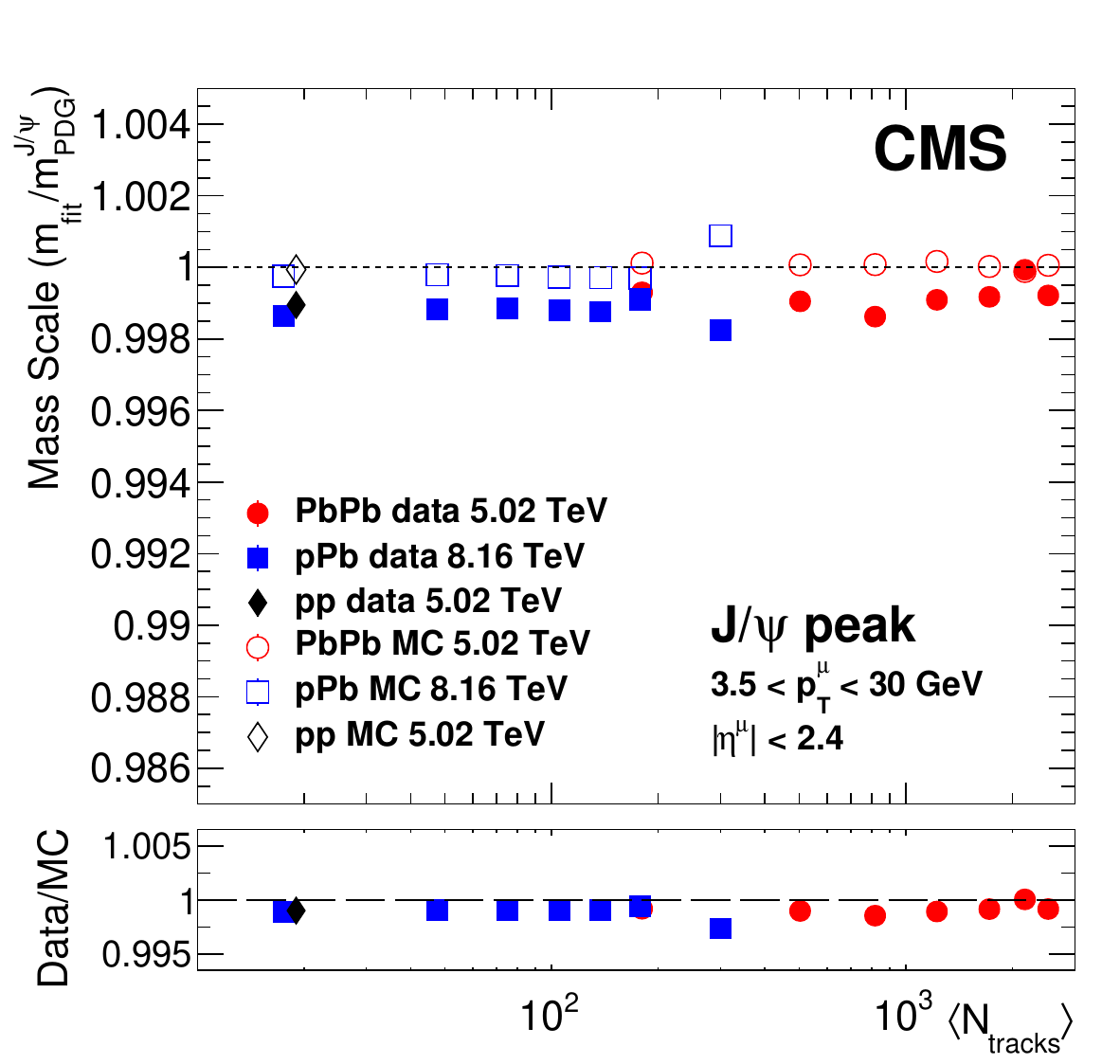}
   \caption{Mass scale at the \Z boson peak (\cmsLeft) and the \JPsi peak (\cmsRight) as a function of the number of tracks from MC simulations (open points) and collision data (filled points) in \pp, \pPb, and \PbPb collisions, calculated as the measured mass \mfit divided by the world-average mass \mPDG. Lower panels show the ratio between data and MC simulation. Only statistical uncertainties are shown.}
\label{fig:MassScale}
\end{figure}

For the \JPsi meson, the BW function is not used because the natural width is negligible. Instead, the signal is described by a sum of a CB and a Gaussian function. In this case, the reconstructed mass, \mfit, is a mean common to the Gaussian core of the CB and to the additional Gaussian function. The two Gaussian width parameters of the signal function are kept distinct to account for the varying resolution in the barrel and endcap regions of the detector. To estimate the mass resolution of the \JPsi peak, we use the average of these two Gaussian widths, weighted by the amount of signal contained in each of the two functions. The extracted width is then divided by the corresponding \mPDG. 

\begin{figure}[htb!]
\centering
\includegraphics[width=\cmsFigWidth]{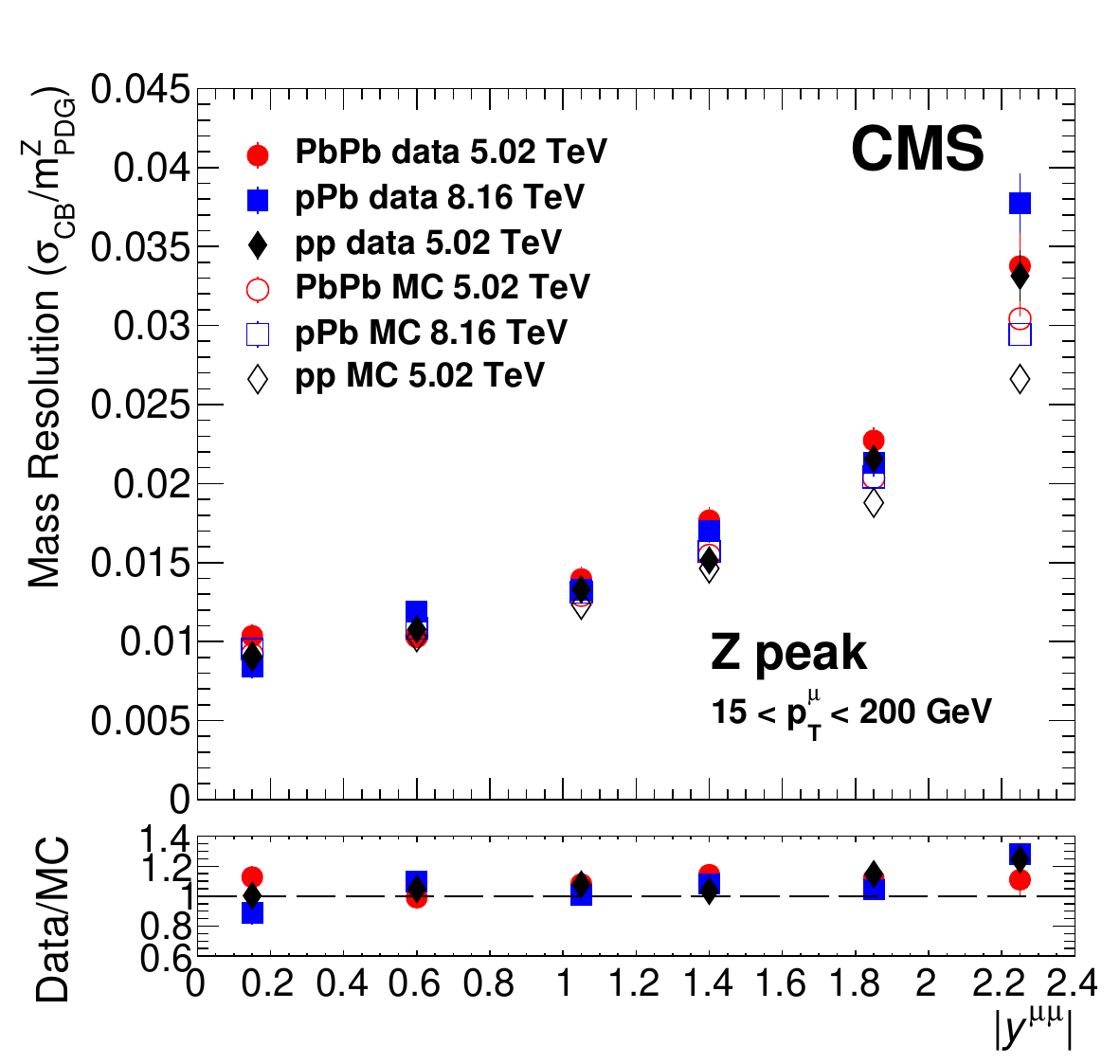}
\includegraphics[width=\cmsFigWidth]{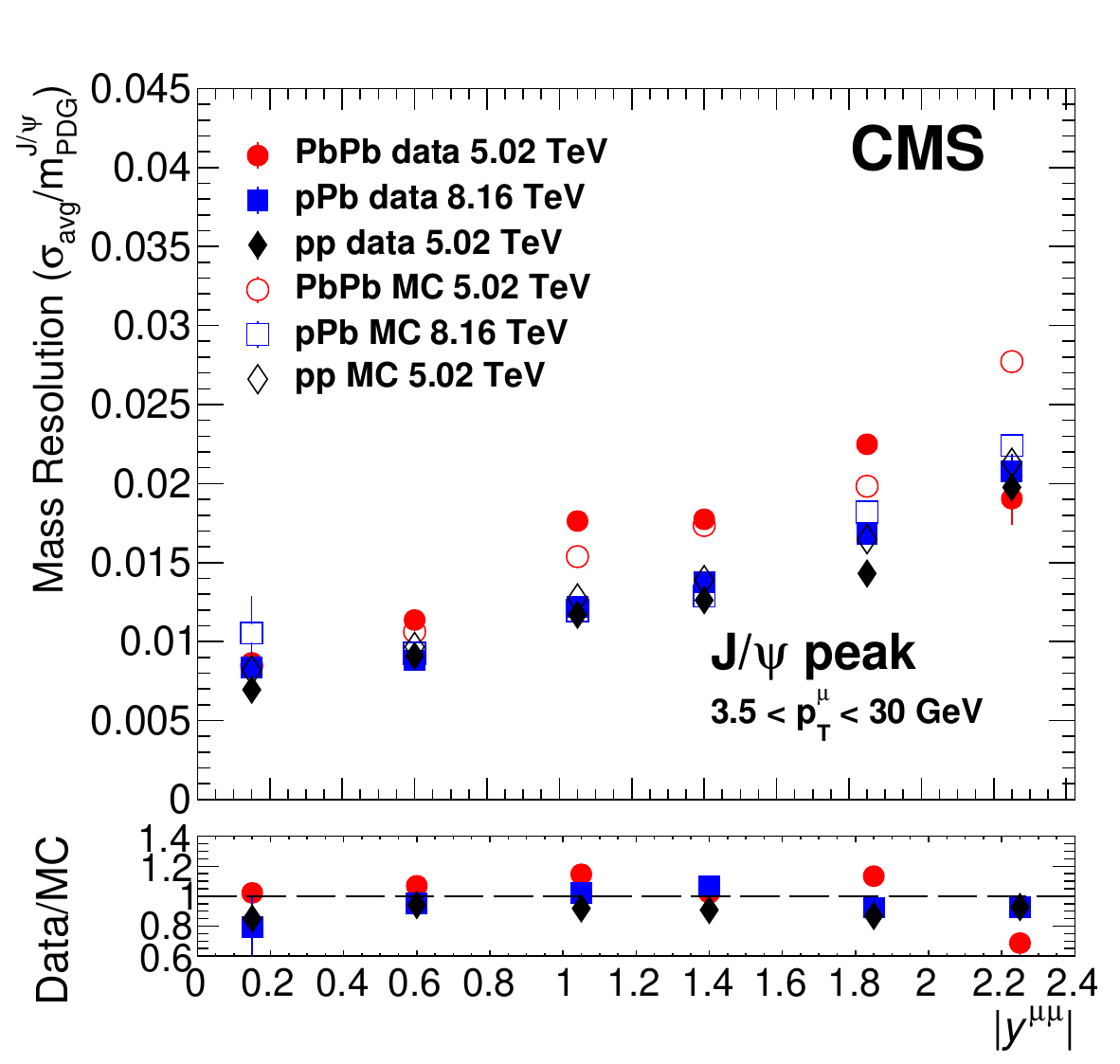}
   \caption{Mass resolution at the \Z boson peak (\cmsLeft) and \JPsi peak (\cmsRight) as a function of $\abs{\dimuonRap}$ from MC simulations (open points) and collision data (filled points) in \pp, \pPb, and \PbPb collisions. The mass resolution is scaled by the world-average mass \mPDG. Lower panels show the ratio between data and MC simulation. Only statistical uncertainties are shown.}
\label{fig:MassResolutionEta}
\end{figure}

\begin{figure}[htb!]
\centering
\includegraphics[width=\cmsFigWidth]{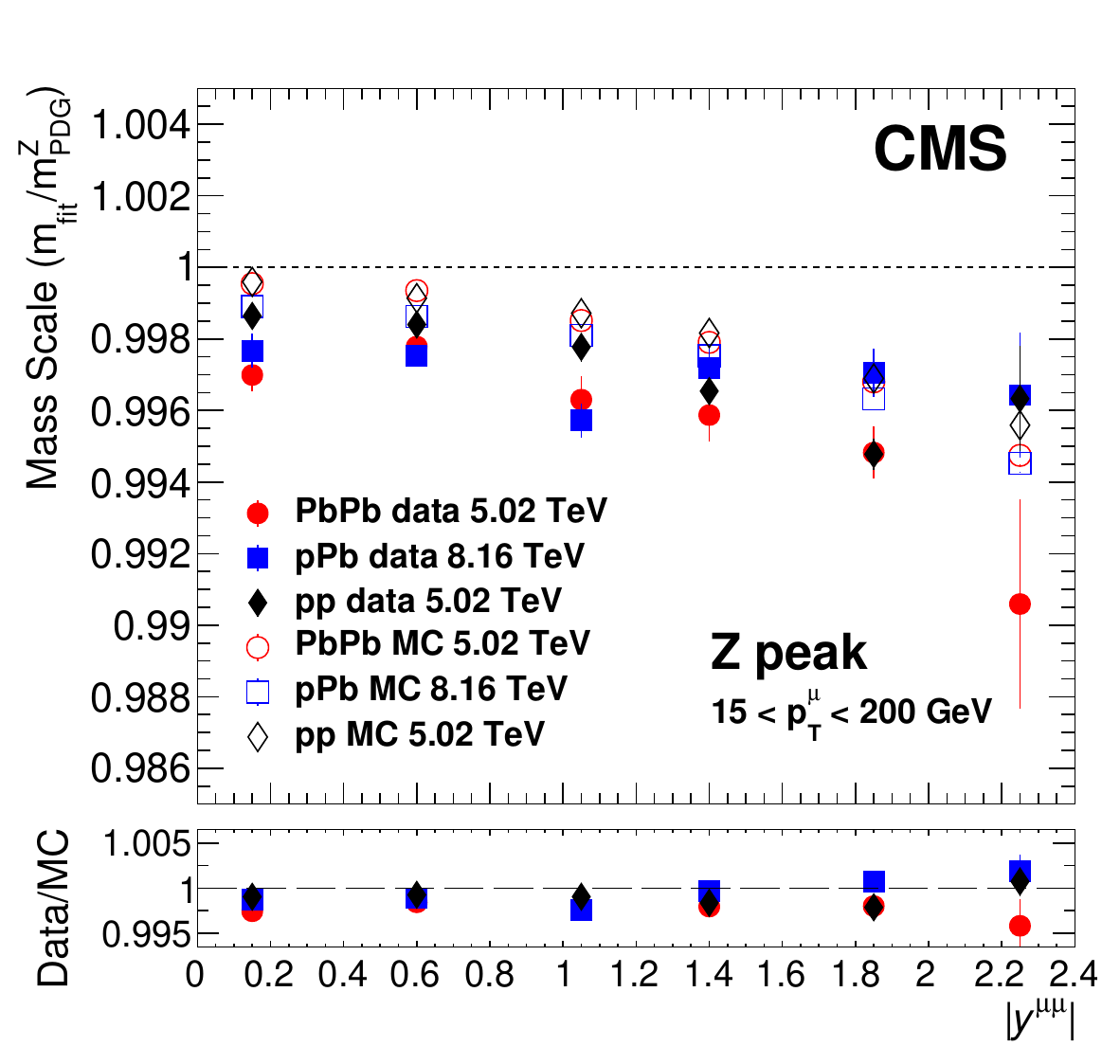}
\includegraphics[width=\cmsFigWidth]{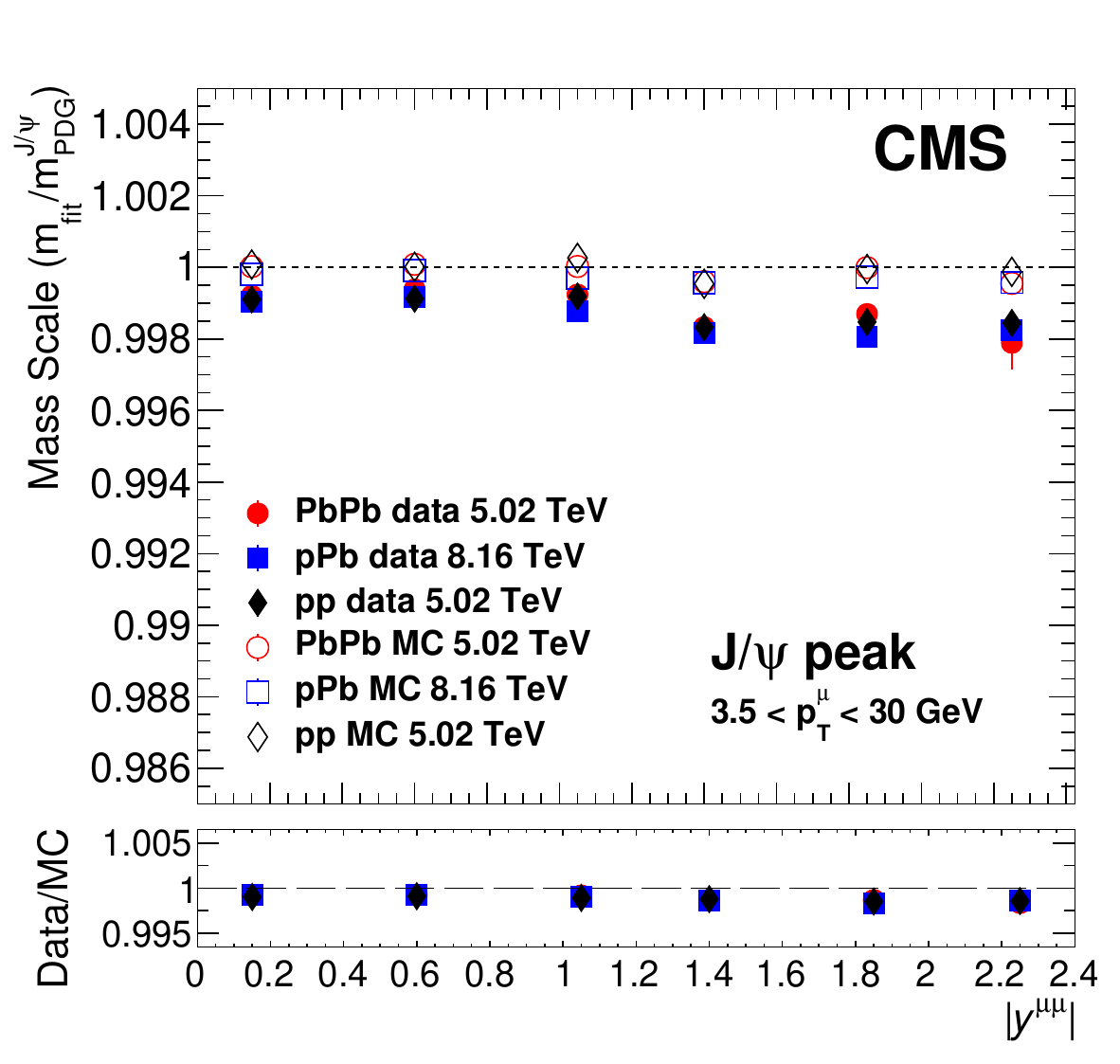}
   \caption{Mass scale at the \Z boson peak (\cmsLeft) and the \JPsi peak (\cmsRight) as a function of $\abs{\dimuonRap}$ from MC simulations (open points) and real data (filled points) in \pp, \pPb, and \PbPb collisions, calculated as the measured mass \mfit divided by the PDG mass \mPDG. Lower panels show the ratio between data and MC simulation. Only statistical uncertainties are shown.}
\label{fig:MassScaleEta}
\end{figure}

The estimated mass resolution is plotted in Fig.~\ref{fig:MassResolution} as a function of \Ntracks for the \Z boson (\cmsLeft) and the \JPsi meson (\cmsRight). The resolution in both cases is consistent with a constant value, which indicates that the resolution is not affected by the level of occupancy in the tracker. The estimated mass scale is plotted in Fig.~\ref{fig:MassScale} as a function of \Ntracks for the \Z boson (\cmsLeft) and the \JPsi meson (\cmsRight). The reconstructed mass is lower than the true mass for both particles, but this scaling is consistent across the different collision systems and for all values of \Ntracks, indicating that the momentum scale is accurate to approximately 0.3\% and is unaffected by the level of occupancy in the tracker or the data-taking year. A change in the mass scale of this order does not generally affect heavy ion measurements, but there are cases (\eg,~Ref.~\cite{Sun:2021joa}) in which such a change can be indicative of interesting physics.

The mass resolution and mass scale versus dimuon rapidity, $\dimuonRap$, are shown in Figs.~\ref{fig:MassResolutionEta} and~\ref{fig:MassScaleEta}, respectively. The resolution varies significantly with rapidity due to the varying angle between the magnetic field $\vec{B}$ and the particle velocity $\vec{v}$. At forward rapidity, where $\vec{v}$ and $\vec{B}$ are almost collinear, a smaller Lorentz force reduces the magnetic bending. The resulting smaller sagitta leads to a greater uncertainty in the momentum measurement. Since the resolution is not constant, for signal description we use the two distinct Gaussian parameters mentioned earlier. The mass scale decreases slightly with rapidity at the \JPsi peak in data but not in MC simulations. At the \Z boson peak, the corresponding decrease is more prominent and is present in both data and MC simulations. In most rapidity bins, the mass scale values tend to be slightly reduced (${\approx}0.2\%$) in data when compared with the MC results. The results are consistent among the \pp, \pPb, and \PbPb collision systems and display the same trends.

\begin{figure}[htb!]
\centering
\includegraphics[width=\cmsFigWidth]{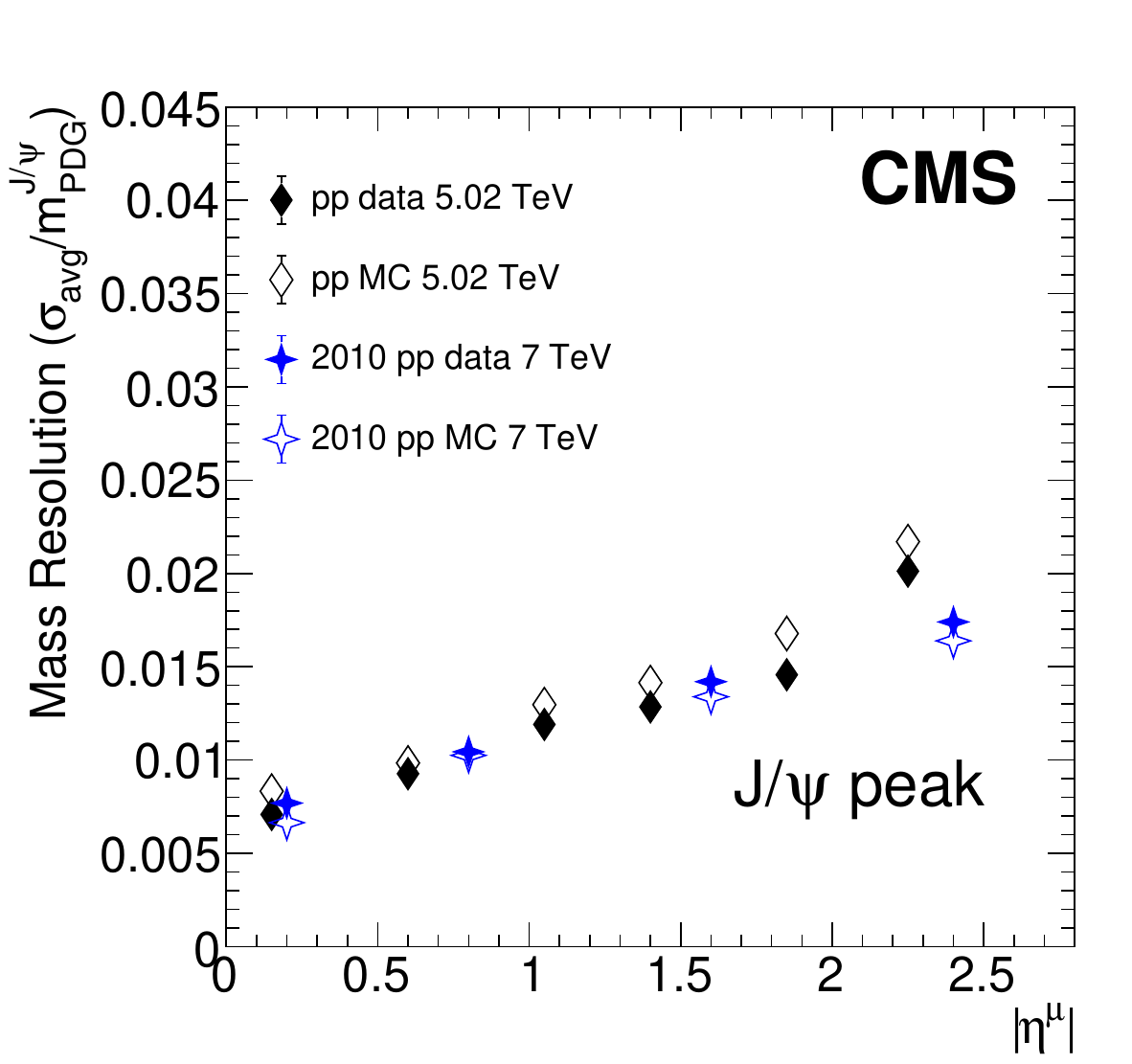}
\includegraphics[width=\cmsFigWidth]{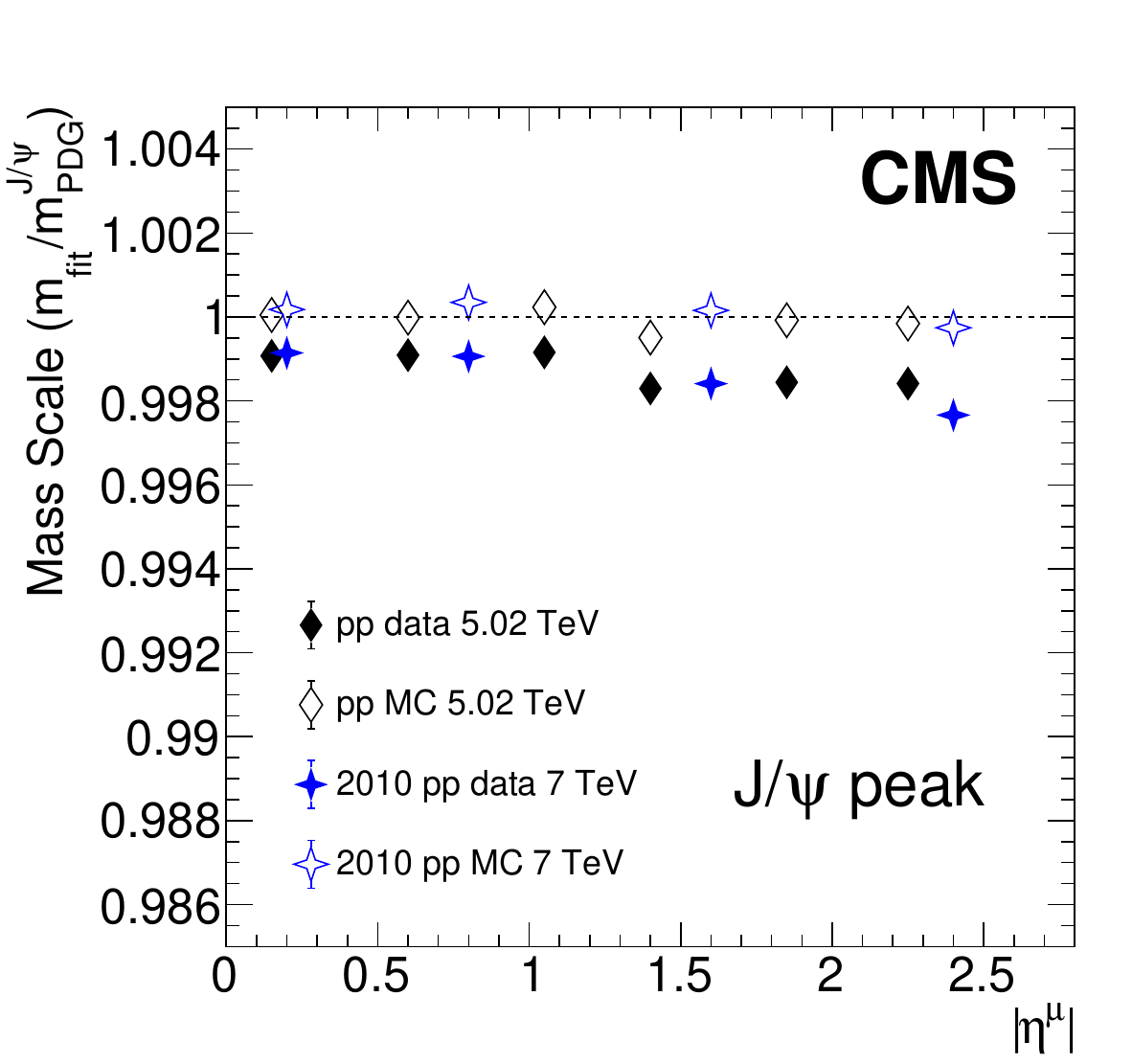}
   \caption{Mass resolution (\cmsLeft) and mass scale (\cmsRight) versus the muon pseudorapidity $\abs{\eta^\PGm}$ at the \JPsi peak in \pp collisions. The values are scaled by the world-average mass $m^{\JPsi}_{\textrm{PDG}}$. The filled points are collision data and the open points are MC simulations. Our results (black diamonds) are compared with measurements from a previous \pp analysis done at $\sqrts=7\TeV$ (blue crosses)~\cite{CMS:2010oua}. Only statistical uncertainties are shown. See details in text.
}
\label{fig:MassResComparison}
\end{figure}

Previous estimates of mass resolution and mass scale were carried out on a \pp data set at a center-of-mass energy $\sqrts=7\TeV$ in 2010~\cite{CMS:2010oua}. A comparison of these results with our data is shown in Fig.~\ref{fig:MassResComparison} as a function of muon pseudorapidity $\abs{\eta^\PGm}$. In order to plot the mass resolution and scale versus pseudorapidity of a single muon, each value of the dimuon mass is entered twice in the distribution used to extract these results: once with the pseudorapidity of the first muon, and then with the pseudorapidity of the second muon. The results of Ref.~\cite{CMS:2010oua} were averaged over the forward and backward pseudorapidity bins. The comparison shows that the muon momentum scale and resolution have been remarkably stable throughout the lifetime of the experiment, indicating that the differences in center-of-mass energy, collision system, accelerator/detector conditions, and kinematic coverage have been successfully addressed by the CMS calibration and reconstruction frameworks.

\section{Summary}
\label{sec:summary}

We have presented efficiencies of muon reconstruction, identification, and triggering, as well as measurements of the dimuon mass scale and resolution of the CMS detector. The efficiencies were estimated using the data-driven tag-and-probe technique. Mass scale and resolution were derived through fits to the invariant mass spectra of \JPsi and \Z boson resonances. We have extended previous studies of the muon performance in \pp collisions to the heavy ion environment using \PbPb data at $\sqrtsNN=5.02\TeV$ and \pPb data at $\sqrtsNN=8.16\TeV$. All the results were also measured in \pp data at $\sqrts=5.02\TeV$ for comparison across all collision systems as a function of the charged particle multiplicity, \Ntracks.

The efficiencies are high (typically above 90\%) in all cases, even at extremely high occupancy. In the low-\pt region, the muon-identification efficiency is comparable between \pp and \PbPb collisions, except in the region of highest occupancy at very low \pt (${<}5\GeV$) and forward pseudorapidity ($\abs{\eta}>2$). This drop in efficiency is expected because the high number of tracks in the inner parts of the detector complicates the matching of tracks between the muon chambers and the tracker. In the high-\pt region, we observe a slight drop (1--2\%) in the muon-identification efficiency in the most central \PbPb events at multiplicities that are unattainable in \pPb or \pp events. A slight decrease (${\approx}3\%$) in the reconstruction efficiency at high occupancies is also observed. Additionally, the trigger efficiency decreases in the most central \PbPb events. A relative reduction of ${\approx}8\%$ in the trigger efficiency occurs between the lowest and highest \Ntracks bins in \PbPb collisions. This reduction is more pronounced than the corresponding decrease in the muon identification and reconstruction efficiencies, suggesting that the CMS single-muon trigger is more sensitive to detector occupancy. In most cases, the efficiencies calculated using Monte Carlo simulations capture the trends seen in data, indicating that the main features are contained in the detector simulation. In a few instances, the MC efficiencies overestimate those obtained from collision data (by up to 4 percentage points), highlighting the need for independent techniques of efficiency estimation based directly on collision data.  

The excellent muon performance of the CMS detector has made possible a robust muon and dimuon program in the heavy ion environment, leading to many muon-based measurements, where some examples include heavy quarkonia, heavy-flavor mesons~\cite{PhysRevLett.107.052302,Chatrchyan_2012,upsilons_2012,psi2s_2014,charmonia_2017,Khachatryan:2016xxp,Khachatryan:2016ypw,UpsilonRatios_2018,charmonia_2018,RAA_Upsilon_2019,v2_Upsilon_2021, CMS:2022sxl}; and electroweak bosons~\cite{W_2012, Zjets_2017, Z_2021, PhysRevLett.106.212301, CMS:2021otx}.

\begin{acknowledgments}
We congratulate our colleagues in the CERN accelerator departments for the excellent performance of the LHC and thank the technical and administrative staffs at CERN and at other CMS institutes for their contributions to the success of the CMS effort. In addition, we gratefully acknowledge the computing centers and personnel of the Worldwide LHC Computing Grid and other centers for delivering so effectively the computing infrastructure essential to our analyses. Finally, we acknowledge the enduring support for the construction and operation of the LHC, the CMS detector, and the supporting computing infrastructure provided by the following funding agencies: SC (Armenia), BMBWF and FWF (Austria); FNRS and FWO (Belgium); CNPq, CAPES, FAPERJ, FAPERGS, and FAPESP (Brazil); MES and BNSF (Bulgaria); CERN; CAS, MoST, and NSFC (China); MINCIENCIAS (Colombia); MSES and CSF (Croatia); RIF (Cyprus); SENESCYT (Ecuador); ERC PRG, RVTT3 and MoER TK202 (Estonia); Academy of Finland, MEC, and HIP (Finland); CEA and CNRS/IN2P3 (France); SRNSF (Georgia); BMBF, DFG, and HGF (Germany); GSRI (Greece); NKFIH (Hungary); DAE and DST (India); IPM (Iran); SFI (Ireland); INFN (Italy); MSIP and NRF (Republic of Korea); MES (Latvia); LMTLT (Lithuania); MOE and UM (Malaysia); BUAP, CINVESTAV, CONACYT, LNS, SEP, and UASLP-FAI (Mexico); MOS (Montenegro); MBIE (New Zealand); PAEC (Pakistan); MES and NSC (Poland); FCT (Portugal); MESTD (Serbia); MCIN/AEI and PCTI (Spain); MOSTR (Sri Lanka); Swiss Funding Agencies (Switzerland); MST (Taipei); MHESI and NSTDA (Thailand); TUBITAK and TENMAK (Turkey); NASU (Ukraine); STFC (United Kingdom); DOE and NSF (USA).

\hyphenation{Rachada-pisek} Individuals have received support from the Marie-Curie program and the European Research Council and Horizon 2020 Grant, contract Nos.\ 675440, 724704, 752730, 758316, 765710, 824093, 101115353, 101002207, and COST Action CA16108 (European Union); the Leventis Foundation; the Alfred P.\ Sloan Foundation; the Alexander von Humboldt Foundation; the Science Committee, project no. 22rl-037 (Armenia); the Belgian Federal Science Policy Office; the Fonds pour la Formation \`a la Recherche dans l'Industrie et dans l'Agriculture (FRIA-Belgium); the Agentschap voor Innovatie door Wetenschap en Technologie (IWT-Belgium); the F.R.S.-FNRS and FWO (Belgium) under the ``Excellence of Science -- EOS" -- be.h project n.\ 30820817; the Beijing Municipal Science \& Technology Commission, No. Z191100007219010 and Fundamental Research Funds for the Central Universities (China); the Ministry of Education, Youth and Sports (MEYS) of the Czech Republic; the Shota Rustaveli National Science Foundation, grant FR-22-985 (Georgia); the Deutsche Forschungsgemeinschaft (DFG), under Germany's Excellence Strategy -- EXC 2121 ``Quantum Universe" -- 390833306, and under project number 400140256 - GRK2497; the Hellenic Foundation for Research and Innovation (HFRI), Project Number 2288 (Greece); the Hungarian Academy of Sciences, the New National Excellence Program - \'UNKP, the NKFIH research grants K 131991, K 133046, K 138136, K 143460, K 143477, K 146913, K 146914, K 147048, 2020-2.2.1-ED-2021-00181, and TKP2021-NKTA-64 (Hungary); the Council of Science and Industrial Research, India; ICSC -- National Research Center for High Performance Computing, Big Data and Quantum Computing and FAIR -- Future Artificial Intelligence Research, funded by the EU NexGeneration program (Italy); the Latvian Council of Science; the Ministry of Education and Science, project no. 2022/WK/14, and the National Science Center, contracts Opus 2021/41/B/ST2/01369 and 2021/43/B/ST2/01552 (Poland); the Funda\c{c}\~ao para a Ci\^encia e a Tecnologia, grant CEECIND/01334/2018 (Portugal); the National Priorities Research Program by Qatar National Research Fund; MCIN/AEI/10.13039/501100011033, ERDF ``a way of making Europe", and the Programa Estatal de Fomento de la Investigaci{\'o}n Cient{\'i}fica y T{\'e}cnica de Excelencia Mar\'{\i}a de Maeztu, grant MDM-2017-0765 and Programa Severo Ochoa del Principado de Asturias (Spain); the Chulalongkorn Academic into Its 2nd Century Project Advancement Project, and the National Science, Research and Innovation Fund via the Program Management Unit for Human Resources \& Institutional Development, Research and Innovation, grant B37G660013 (Thailand); the Kavli Foundation; the Nvidia Corporation; the SuperMicro Corporation; the Welch Foundation, contract C-1845; and the Weston Havens Foundation (USA).
\end{acknowledgments}

\bibliography{auto_generated}
\numberwithin{figure}{section}
\appendix

\section{Converting centrality to number of tracks}
\label{Appendix:CentralityToNtracks}

It is common in analyses of \PbPb collisions to study observables in bins of centrality, whereas in \pp and \pPb collisions it is common to do so in bins of \Ntracks. In order to compare the existing efficiency results in all three collision systems as a function of a common event activity variable, we converted the \PbPb centrality bins to corresponding average values of \Ntracks, which we label $\langle \Ntracks \rangle$ in tables and figures. The \Ntracks variable was chosen because it is suitable for studying occupancy-related effects.
 
\begin{figure}[htb]
\centering
\includegraphics[width=\cmsFigWidth]{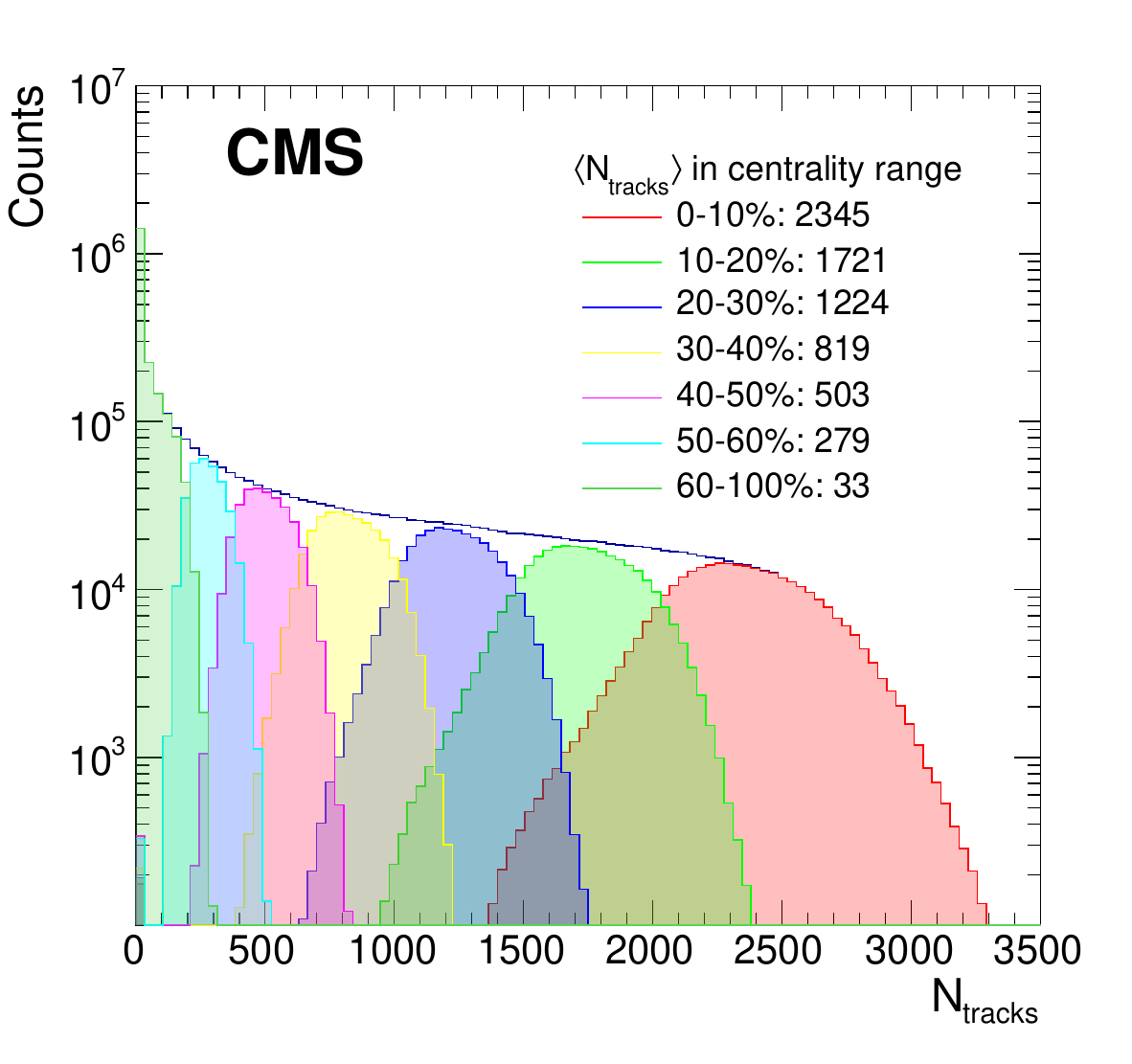}
\includegraphics[width=\cmsFigWidth]{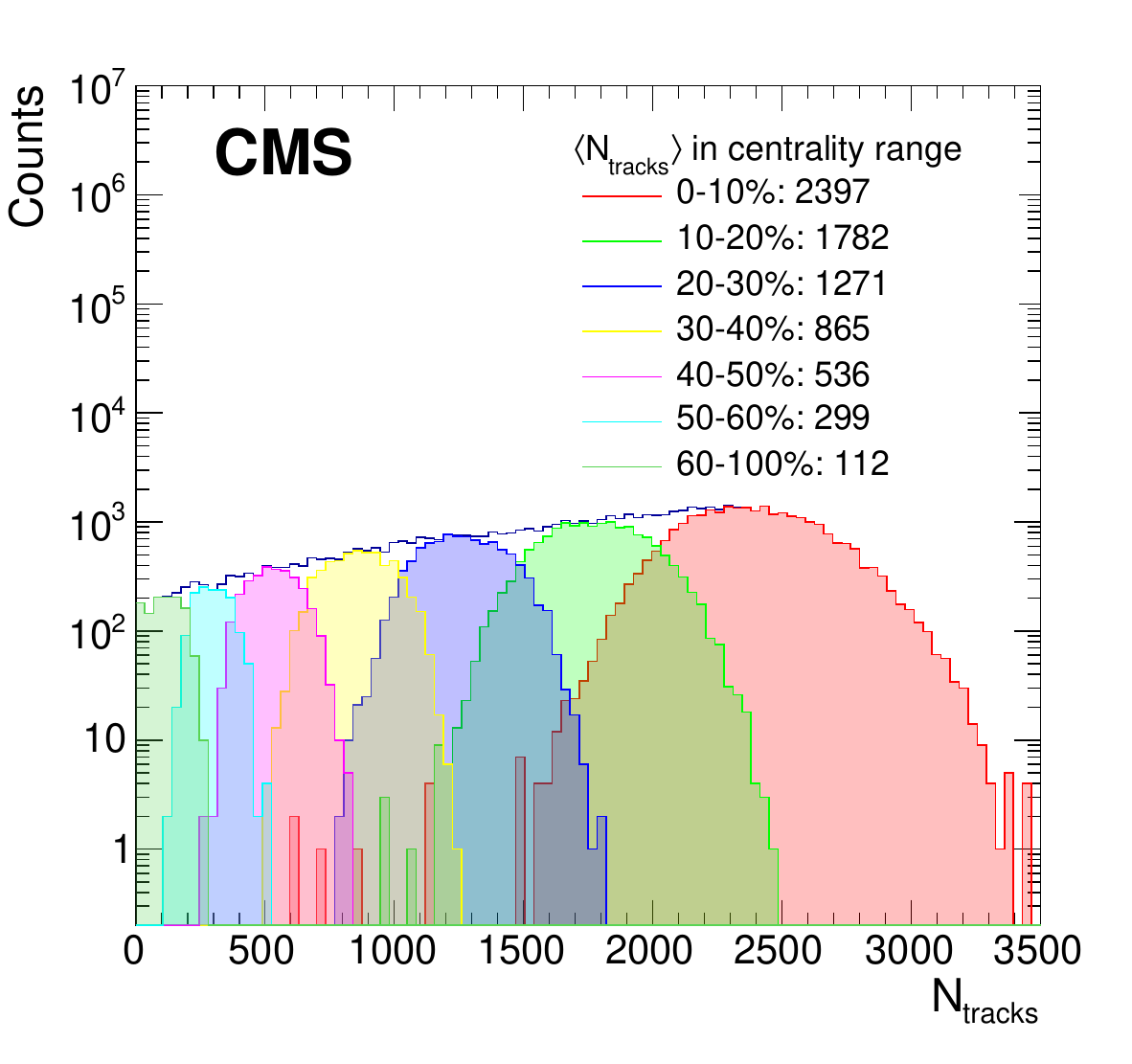}
\caption{Distributions of \Ntracks in the \PbPb collisions for various ranges of centrality in a minimum-bias (\cmsLeft) and a muon-triggered (\cmsRight) data set.}
\label{fig:centToNtracks}
\end{figure}

The conversion was done by plotting the distribution of \Ntracks in the \PbPb data set for each centrality bin and taking the mean of this distribution as the value of \Ntracks for that efficiency datum point. Examples of the conversion procedure are shown in Fig.~\ref{fig:centToNtracks}. The plot on the \cmsLeft was generated from a minimum-bias data set, which required that events must have a coincidence signal in both the positive-$z$ and negative-$z$ side of the HF. The plot on the \cmsRight was from a muon-triggered data set, which required that events must have at least two muons passing various selection criteria common to dimuon analyses. Requiring two muons biases the average \Ntracks toward higher values. The triggered data are used to translate centrality bins to \Ntracks because they describe most accurately the occupancy conditions of the detector for our data samples.

\cleardoublepage \section{The CMS Collaboration \label{app:collab}}\begin{sloppypar}\hyphenpenalty=5000\widowpenalty=500\clubpenalty=5000\input{MUO-21-001-public-authorlist.tex}\end{sloppypar}
\end{document}

%% file: MUO-21-001-public-authorlist.tex
\cmsinstitute{Yerevan Physics Institute, Yerevan, Armenia}
{\tolerance=6000
A.~Hayrapetyan, A.~Tumasyan\cmsAuthorMark{1}\cmsorcid{0009-0000-0684-6742}
\par}
\cmsinstitute{Institut f\"{u}r Hochenergiephysik, Vienna, Austria}
{\tolerance=6000
W.~Adam\cmsorcid{0000-0001-9099-4341}, J.W.~Andrejkovic, T.~Bergauer\cmsorcid{0000-0002-5786-0293}, S.~Chatterjee\cmsorcid{0000-0003-2660-0349}, K.~Damanakis\cmsorcid{0000-0001-5389-2872}, M.~Dragicevic\cmsorcid{0000-0003-1967-6783}, P.S.~Hussain\cmsorcid{0000-0002-4825-5278}, M.~Jeitler\cmsAuthorMark{2}\cmsorcid{0000-0002-5141-9560}, N.~Krammer\cmsorcid{0000-0002-0548-0985}, A.~Li\cmsorcid{0000-0002-4547-116X}, D.~Liko\cmsorcid{0000-0002-3380-473X}, I.~Mikulec\cmsorcid{0000-0003-0385-2746}, J.~Schieck\cmsAuthorMark{2}\cmsorcid{0000-0002-1058-8093}, R.~Sch\"{o}fbeck\cmsorcid{0000-0002-2332-8784}, D.~Schwarz\cmsorcid{0000-0002-3821-7331}, M.~Sonawane\cmsorcid{0000-0003-0510-7010}, S.~Templ\cmsorcid{0000-0003-3137-5692}, W.~Waltenberger\cmsorcid{0000-0002-6215-7228}, C.-E.~Wulz\cmsAuthorMark{2}\cmsorcid{0000-0001-9226-5812}
% ...
\par}
\vskip\cmsinstskip
\dag:~Deceased\\
$^{1}$Also at Yerevan State University, Yerevan, Armenia\\
$^{2}$Also at TU Wien, Vienna, Austria\\
% ...
